\documentclass[aps,twocolumn,superscriptaddress,groupedaddress,prx]{revtex4-2}
\usepackage{adjustbox}
\usepackage{dcolumn}   
\usepackage{bm,amsmath,amssymb}
\usepackage{graphicx}
\usepackage{subfigure}
\usepackage[dvipsnames]{xcolor}
\usepackage{mathtools}
\usepackage{amsthm}
\usepackage{braket}
\usepackage[english]{babel}
\usepackage[utf8]{inputenc}
\usepackage[colorlinks=true]{hyperref}
\usepackage{color}
\definecolor{Ora}{cmyk}{0, 0.6, 0.8, 0}


\graphicspath{{new_figures/}}

\def\vecb#1{\boldsymbol{#1}}
\def\ket#1{|#1\rangle}

\def\scal#1#2{\langle#1|#2\rangle}
\def\matr#1#2#3{\langle#1|#2|#3\rangle}

\def\ave#1{\langle#1\rangle}

\def\Lambi{\Lambda_{\rm I}}
\def\Lambf{\Lambda_{\rm F}}
\def\bLambi{\vecb{\Lambda}_{\rm I}}
\def\bLambf{\vecb{\Lambda}_{\rm F}}
\def\bLambda{\vecb{\Lambda}}
\def\psif{\psi_{\rm F}}

\def\uvo#1{\lq\lq #1\rq\rq}

\def\smallunderbrace#1{\mathop{\vtop{\m@th\ialign{##\crcr
   $\hfil\displaystyle{#1}\hfil$\crcr
   \noalign{\kern3\p@\nointerlineskip}%
   \tiny\upbracefill\crcr\noalign{\kern3\p@}}}}\limits}

\begin{document}

\title{Search for optimal driving in finite quantum systems with precursors of criticality}

	\author{Felipe~Matus}
	\email{matus@ipnp.mff.cuni.cz}
	\affiliation{Institute of Particle and Nuclear Physics, Faculty of Mathematics and Physics, Charles University,
		V Hole{\v s}ovi{\v c}k{\' a}ch 2, 180 00 Prague, Czechia}
	\author{Jan St{\v r}ele{\v c}ek}
	\email{strelecek@ipnp.mff.cuni.cz}
	\affiliation{Institute of Particle and Nuclear Physics, Faculty of Mathematics and Physics, Charles University,
		V Hole{\v s}ovi{\v c}k{\' a}ch 2, 180 00 Prague, Czechia}
	\author{Pavel Str{\' a}nsk{\' y}} 
	\email{stransky@ipnp.mff.cuni.cz}
	\affiliation{Institute of Particle and Nuclear Physics, Faculty of Mathematics and Physics, Charles University,
		V Hole{\v s}ovi{\v c}k{\' a}ch 2, 180 00 Prague, Czechia}
	\author{Pavel Cejnar}
	\email{cejnar@ipnp.mff.cuni.cz}
	\affiliation{Institute of Particle and Nuclear Physics, Faculty of Mathematics and Physics, Charles University,
		V Hole{\v s}ovi{\v c}k{\' a}ch 2, 180 00 Prague, Czechia}

	\date{\today}

\begin{abstract}
Using the adiabatic perturbation theory of driven dynamics [Phys.\,Rev.\,A~{\bf 78},\,052508\,(2008)] we design a hierarchy of quantum state preparation protocols that systematically increase the fidelity at very long driving times.
We test these and other protocols, including those based on the geometric analysis of the parameter space, in a~single-qubit system and in a fully connected multi-qubit system showing in its infinite-size limit several quantum phase transitions. 
The new protocols excel in the asymptotic driving regime, above a~crossover time from the Landau-Zener regime which increases with a decreasing minimal energy gap along the driving path (with the size of the system).
In the medium-time domain, the performance of all tested protocols is indecisive.
\end{abstract}

\maketitle

\section{INTRODUCTION}
\label{sec:intro}

One of the big challenges of modern physics and technology is to build a scalable quantum computer.
The idea of so-called adiabatic quantum computation \cite{Fahr00,Alba18} is based on encoding a particular computational problem into complex correlations involved in a quantum state of an interacting many-body system.
The most commonly discussed method of a noiseless preparation of such a state relies on the adiabatic theorem of quantum mechanics \cite{Born28,Kato50,Mess62,Saku11}.
The system is initially prepared in an easily obtainable, uncorrelated configuration, such as a fully oriented state of a spin lattice.
This state represents the ground state of the lattice in a strong external magnetic field and can be prepared by cooling down the system with field to nearly zero absolute temperature.
In contrast, the desired highly correlated state reflects mutual interactions between individual spins in the ground state of the lattice in absence of the external field.
This state cannot be efficiently produced by cooling down the system without field (the free energy landscape is assumed to have numerous local minima), but can be obtained from the uncorrelated initial configuration by a very slow, nearly adiabatic attenuation of the field. 
This in an ideal case prevents any excitation of the lattice, keeping it in the instantaneous ground state up to the moment when the field completely vanishes.

Cast in a general form, the above method of adiabatic preparation of target states represents an externally driven slow change of some control parameters which keeps the system in a selected discrete eigenstate (the ground state) of the evolving Hamiltonian.
The method can be used not only for purely quantum computational purposes, but also in other related techniques of quantum information processing \cite{Guer19,Schi22}.
However, to realize such a driving protocol in practice is usually a rather difficult task because finite-time corrections to the adiabatic evolution can be rather large \cite{Garr62,Berr87,Nenc93,Teuf03,Orti08,Orti10,Orti14}.
Obstacles to adiabaticity may follow from various system-specific structural properties. 
In particular, the probability of unwanted excitations of the system increases as a consequence of accidental or systematic enlargement of transition matrix elements or reduction of energy gaps between individual levels in some parameter regions. 
To perform the state preparation protocol with a high fidelity requires a slow down of the parameter change in these regions, which sets lower bounds on the minimal time required. 
This is tightly connected with latterly widely discussed topic of quantum speed limits \cite{Deff17,Fran16,Buko19}.

The adiabatic state preparation techniques are most problematic if the uncorrelated and correlated initial and final states belong to different quantum phases of the system, being therefore separated by a finite-size precursor of a quantum phase transition (QPT) \cite{Sach99,Carr10}.
The energy gap between the ground state and the first excited state at the critical point converges to zero with an increasing size of the system, which may lead to the loss of scalability of the protocol---a too rapid increase of time with size~\cite{Schu06}.
Various aspects of driving through a~QPT, including links to the celebrated Kibble-Zurek mechanism, were studied, e.g., in Refs.\,\cite{Dams05,Zure05,Polk05,Fubi07,Dzia10,Mish18,Hart19,Sinh21}. 

The problem of high-fidelity driving in quantum precritical system is addressed in this paper.
We combine an analytic approach based on the adiabatic perturbation theory (APT) in the form of Ref.\,\cite{Orti08} with numerical simulations of driven dynamics in systems composed of ${N\geq 1}$ fully connected qubits.
We show that previously discussed driving protocols based on the geometric approach (using the geodesic path in the parameter space according to the Provost-Vallee metric) \cite{Prov80,Tomk16,Kolo17,Buko19} do not give (in general) the best results.
We nevertheless propose a class of driving protocols that yield increasingly high fidelity for very long driving times, in the regime where the APT dominates.
We show that the transition to this regime from the Landau-Zener regime, which dominates at smaller times, has a character of a sharp crossover.
It takes place at times that grow with a decreasing minimal energy gap on the driving trajectory, i.e., with an increasing size of the system if the minimal gap coincides with crossing of the QPT.  

The plan of the paper is as follows:
In Sec.\,\ref{sec:apt} we outline the APT of Ref.\,\cite{Orti08} and describe its application to the driving problem, designing the above-mentioned new class of driving protocols. 
These are subsequently compared with several other protocols, all summarized in Sec.\,\ref{sec:drivings}, including the protocols based on the geometric approach.
Results of numerical simulations of driven dynamics in systems composed of one or more qubits are described in Secs.\,\ref{se:two} and~\ref{se:many}, respectively.
The one-qubit system with an avoided crossing serves as a treatable toy model for a more complex behavior observed in an interacting multi-qubit system with finite-size precursors of QPTs of various kinds.
The simulations demonstrate a high fidelity obtained in the newly proposed APT-based protocols in sufficiently long times and contest the alleged general supremacy of protocols based on the geometric approach.
Brief summary and conclusions come in Sec.\,\ref{CONC}. 

Note that in this paper we set ${\hbar = 1}$.
Decadic and natural logarithms are distinguished by symbols $\log$ and $\ln$, respectively.

\section{ADIABATIC PERTURBATION THEORY}
\label{sec:apt}

Various perturbative approaches to the dynamics of slowly driven systems have been discussed in the literature---see, e.g., Refs.\,\cite{Garr62,Berr87,Nenc93,Teuf03,Orti08,Orti10,Orti14}.
In this work we use the APT developed by Rigolin, Ortiz and Ponce~\cite{Orti08,Orti10,Orti14}.
The present section briefly outlines the theory and shows its application to the maximum fidelity problem.

\subsection{Outline of the theory}
\label{sec:outline}

We consider a quantum system with $d$-dimensional Hilbert space ${\cal H}$ (the dimension can also be infinite) and Hamiltonian $\hat{H}(\bLambda)$ depending on a finite set of real control parameters $\bLambda\equiv{(\Lambda^1,\Lambda^2,\ldots,\Lambda^D)}$ (external fields and/or coupling constants) that form a $D$-dimensional parameter space.
Eigenvalues and the corresponding eigenvectors of $\hat{H}(\bLambda)$ are denoted as $E_n(\bLambda)$ and $\ket{E_n(\bLambda)}$, respectively.
In accord with Ref.\,\cite{Orti08}, we assume a~fully nondegenerate spectrum in the parameter region relevant for the driving, which implies unique identification of eigenvectors at each point~$\bLambda$.
The eigenvalues are ordered in an increasing manner with ${n=0}$ corresponding to the ground state. 
Energy differences are denoted by
\begin{equation}\label{Del}
\Delta_{nm}(\bLambda) = E_n(\bLambda)-E_m(\bLambda).
\end{equation}

The Hamiltonian parameters $\bLambda$ are varied in a prescribed way, following a path 
\begin{equation}\label{path}
{\wp}\equiv\left\{\bLambda(t)=\bigl(\Lambda^{1}(t), \Lambda^{2}(t), \ldots, \Lambda^{D}(t)\bigr)\right\}_{t=0}^{T},
\end{equation}
which starts at ${\bLambda(0)=\bLambi}$ at the initial time ${t=0}$ and ends at ${\bLambda(T)=\bLambf}$ at the final time ${t=T}$.
Let us stress that Eq.\,\eqref{path} defines not only the geometric shape of the path in the parameter space, but also speeds, accelerations and all higher derivatives at all points along the path. 
Defining a rescaled time
\begin{equation} 
\tau = \frac{t}{T}\in[0,1],
\end{equation}
we obtain a parametrization $\bLambda(\tau)$ of the path independent of the total driving time $T$.
All derivatives $\frac{d^k}{dt^k}\Lambda^{\mu}(t)$ scale with the respective power of $T$. 
In the following, the derivatives with respect to $\tau$ will be denoted by dots, so for instance ${\dot{\Lambda}^{\mu}(\tau)\equiv\frac{d}{d\tau}\Lambda^{\mu}(\tau)=T\frac{d}{dt}\Lambda^{\mu}(t)}$, ${\ddot{\Lambda}^{\mu}(\tau)\equiv\frac{d^2}{d\tau^2}\Lambda^{\mu}(\tau)=T^2\frac{d^2}{dt^2}\Lambda^{\mu}(t)}$ and so on.

Our task is to find the evolution of the state vector $\ket{\psi(\tau)}_{\wp}$ induced by the time-dependent Hamiltonian ${\hat{H}(\bLambda(\tau))\equiv\hat{H}(\tau)}$ associated with a general parameter path~$\wp$.
In particular, starting from an initial state $\ket{\psi(0)}_{\wp}$, we want to determine the overlap of the final state $\ket{\psi({1})}_{\wp}$ with a chosen state $\ket{\psif}$ to be prepared.
The usual choice, applied also in this work, is 
\begin{eqnarray}
\ket{\psi(0)}_{\wp}&=&\ket{E_0(\bLambi)},
\label{inic}
\\
\ket{\psif}&=&\ket{E_0(\bLambf)}.
\label{finst}
\end{eqnarray}
In the APT, the exact solution to the Schr{\"o}dinger equation is searched as an expansion in powers of $T^{-1}$, 
\begin{equation}\label{series}
\ket{\psi(\tau)}_{\wp} =\lim_{P\to\infty} {\cal N}_P(\tau)_{\wp}\sum_{p=0}^{P}T^{-p}\ket{\psi^{(p)}(\tau)}_{\wp},
\end{equation}
where $\ket{\psi^{(p)}(\tau)}_{\wp}$ is the $p$th-order correction of the state vector for the particular path $\wp$ and 
\begin{equation}
{\cal N}_P(\tau)_{\wp}=\biggl[\ \sum_{p,p'=1}^{P}T^{-(p+p')}\scal{\psi^{(p)}(\tau)}{\psi^{(p')}(\tau)}_{\wp}\biggr]^{-\frac{1}{2}}
\end{equation}
is a normalization coefficient of the expansion up to the order ${p=P}$.
The highest order $P$ goes to infinity in the exact solution \eqref{series}, but it can be set to a finite value to get a reasonable approximation of the exact solution for sufficiently large $T$.
Here and in the following, all entities with subscript $\wp$ depend on the specific path~\eqref{path}, while those depending only on $\tau$ can be determined from the local properties of the system at $\bLambda(\tau)$ and from the instantaneous speed $\dot{\bLambda}(\tau)$.   
We note that the assignment to ${\wp}$ brings a certain residual dependence on the perturbation parameter $T^{-1}$ into the expansion \uvo{coefficients} $\ket{\psi^{(p)}(\tau)}_{\wp}$, which is in contrast to usual perturbation techniques. 

The $p$th term in the series \eqref{series} can be expressed in the eigenbasis $\ket{E_n(\bLambda(\tau))}\equiv\ket{E_n(\tau)}$ of the instantaneous Hamiltonian,
\begin{equation}\label{psip}
\ket{\psi^{(p)}(\tau)}_{\wp} = \sum_{n = 0}^{d-1} e^{-i\varphi_n(\tau)_{\wp}}\ b_{n}^{(p)}(\tau)_{\wp}\ \ket{E_n(\tau)},
\end{equation}
where $b_{n}^{(p)}(\tau)_{\wp}$ are expansion coefficients and
\begin{equation}\label{phase}
\varphi_n(\tau)_{\wp}=\omega_{n}(\tau)_{\wp}T-\gamma_{n}(\tau)_{\wp}
\end{equation}
are phases, each composed of the dynamical phase $\omega_{n}(\tau)_{\wp}T$ and the geometrical phase $\gamma_{n}(\tau)_{\wp}$.
We have
\begin{eqnarray}
	\omega_{n}(\tau)_{\wp} &=&\int_{0}^{\tau} E_n(\tau')\,d\tau',
	\label{dyna}
	\\
	\gamma_{n}(\tau)_{\wp} &=& i\int_{0}^{\tau}M_{nn}(\tau')\,d\tau',
	\label{geom}
\end{eqnarray}
where we employed diagonal elements (which can be proven to be pure imaginary) of the matrix
\begin{eqnarray}\label{Mnm}
M_{nm}(\tau)&=&\bigl\langle E_n(\tau) \bigr| \tfrac{d}{d\tau}E_m (\tau) \bigr\rangle
\\
&=&\begin{dcases}
\dot{\Lambda}^{\mu}(\tau)\bigl\langle E_n(\tau) \bigr| \tfrac{\partial}{\partial\Lambda^{\mu}}E_n (\tau) \bigr\rangle & n=m,
\\
-\dot{\Lambda}^{\mu}(\tau) \frac{
 \bigl\langle E_{n}(\tau)\bigr|\frac{\partial}{\partial\Lambda^{\mu}}\hat{H}(\tau)\bigl| E_{m}(\tau)\bigr\rangle}
 {\Delta_{nm}(\tau)} & n\neq m.
\end{dcases}
\nonumber
\end{eqnarray}
The Einstein summation convention is used for index $\mu$ (this convention will be kept for Greek indices everywhere below) and
the energy difference \eqref{Del} is evaluated at ${\bLambda=\bLambda(\tau)}$.
We stress that the explicit separation of phases \eqref{dyna} and \eqref{geom} in Eq.\,\eqref{psip} is useful since for slow driving near the adiabatic limit these represent the principal contributions to the overall phase.
The zeroth-order contribution to the expansion \eqref{series} for the initial condition~\eqref{inic} is set to coincide with the adiabatic solution
\begin{equation}\label{zero}
\ket{\psi^{(0)}(\tau)}_{\wp}=e^{-i\varphi_0(\tau)_{\wp}}\ket{E_0(\tau)}.
\end{equation}
We note that the phases of eigenvectors $\ket{E_n(\tau)}$, which enter through Eq.\,\eqref{Mnm} into the APT formulas below, are fixed by the requirement of continuity of eigenvectors along the driving path.
At the initial point ${\tau=0}$, e.g., the phases can be chosen arbitrarily.

To obtain an iterable expression of the evolving state vector \eqref{series}, the expansion coefficients in Eq.\,\eqref{psip} are further expanded as 
\begin{equation}\label{bn}
	b_{n}^{(p)}(\tau)_{\wp} = \sum_{m=0}^{d-1} e^{i\varphi_{nm}(\tau)_{\wp}}\ b_{nm}^{(p)}(\tau)_{\wp},
\end{equation}
where $b_{nm}^{(p)}(\tau)_{\wp}$ are new coefficients and
\begin{equation}\label{phadi}
\varphi_{nm}(\tau)_{\wp} = \varphi_n(\tau)_{\wp} - \varphi_m(\tau)_{\wp}
\end{equation}
are phase differences.
The expansion \eqref{bn} may seem redundant, being just a re-expression of each coefficient $b_n^{(p)}(\tau)_{\wp}$ in terms of many new coefficients $b_{nm}^{(p)}(\tau)_{\wp}$, $m=0,1,\ldots,{d-1}$. 
Nevertheless, this ansatz plays a crucial role in the formulation of the APT in Ref.\,\cite{Orti08} as the new coefficients satisfy a recurrent formula
\begin{equation}\label{iter}
\begin{array}{l}
i\Delta_{nm}(\tau)\ b_{nm}^{(p+1)}(\tau)_{\wp}+\dot{b}_{nm}^{(p)}(\tau)_{\wp}
\\
\quad+\bigl[M_{nn}(\tau)-M_{mm}(\tau)\bigr]\ b_{nm}^{(p)}(\tau)_{\wp}
\\
\qquad+\sum\limits_{k(\neq n)}M_{nk}(\tau)\,b_{km}^{(p)}(\tau)_{\wp}=0,
\end{array}
\end{equation}
which contains only snapshots of the quantities involved at specific time $\tau$ and allows for an iterative solution. 
Although the resulting expressions for $b_{nm}^{(p)}(\tau)_{\wp}$ acquire, in general, the path dependence via some integrals over ${\tau'\in[0,\tau]}$ (see below), the locality of the condition \eqref{iter} is very suitable for its practical solution.

As already pointed out, solving of Eq.\,\eqref{iter} proceeds in an iterative way, so the coefficients $b_{nm}^{(p+1)}(\tau)$ are determined from $b_{nm}^{(p)}(\tau)$.
For the initial condition \eqref{inic} we start the iteration from ${b_{nm}^{(0)}(\tau)=\delta_{n0}\delta_{m0}}$, which is equivalent to the adiabatic ansatz for the ${p=0}$ term in Eq.\,\eqref{zero}.
Details of the recursive determination of general ${p=1}$ and ${p=2}$ terms are described in Ref.\,\cite{Orti08}.
Here we explicitly show only the ${p=1}$ term,
\begin{equation}\label{bn1}
b_{n}^{(1)}(\tau)_{\wp}\!=\!
\begin{dcases}
i\sum\limits_{m=1}^{d-1}\int_0^{\tau}\frac{|M_{m0}(\tau')|^2}{\Delta_{m0}(\tau')}d\tau' & n\!=\!0,
\\
i\left[ e^{i\varphi_{n0}(\tau)_{\wp}}\frac{M_{n0}(\tau)}{\Delta_{n0}(\tau)}\!-\!\frac{M_{n0}(0)}{\Delta_{n0}(0)}\right] & n\!>\!0,
\end{dcases}
\qquad
\end{equation}
where we use off-diagonal elements of the matrix \eqref{Mnm}, energy differences \eqref{Del} and the phase differences \eqref{phadi}.
We stress that expression \eqref{bn1} is valid only for the initial condition \eqref{inic}.

\subsection{Application to the maximum fidelity problem}
\label{sub:apliAPT}

The overlap of the final ${t=T}$ state $\ket{\psi({\tau=1})}_{\wp}$ of the system with the target state $\ket{\psif}$ is characterized by the fidelity ${\cal F}({\tau=1})_{\wp}=\bigl|\scal{\psif}{\psi({\tau=1})}_{\wp}\bigr|^2$ (simply the probability of identifying the evolved state with the target one).
Its value between 0 (no overlap) and 1 (full overlap) quantifies the success of the completed driving protocol.
Since $\ket{\psif}$ coincides with the ground state $\ket{E_0(\bLambf)}$ of the final Hamiltonian, see Eq.\,\eqref{finst}, it is convenient to measure an overlap of the evolving state with the instantaneous ground state: 
\begin{equation}\label{fidel}
{\cal F}(\tau)_{\wp}=\left|\scal{E_0(\tau)}{\psi(\tau)}_{\wp}\right|^2=1-{\cal I}(\tau)_{\wp}.
\end{equation}
This provides an evolving fidelity value which converges to the resulting fidelity at ${\tau=1}$.
The time dependence of ${\cal F}(\tau)_{\wp}$ will help us to monitor the progress of the state preparation protocol, e.g., to identify the parameter domains where the system is easily excitable.
The complementary time-dependent quantity ${\cal I}(\tau)_{\wp}= 1-{\cal F}(\tau)_{\wp}$ is named infidelity.

Following the formalism of Sec.\,\ref{sec:outline}, we can expand the evolving fidelity in powers of $1/T$,
\begin{eqnarray}\label{seriesF}
{\cal F}(\tau)_{\wp}&=&
\lim_{P\to\infty}\biggl|{\cal N}_P(\tau)_{\wp}\sum_{p=0}^{P}T^{-p}\scal{E_0(\tau)}{\psi^{(p)}(\tau)}_{\wp}\biggr|^2
\nonumber\\
&=&\sum_{p=0}^{\infty}T^{-p}{\cal F}^{(p)}(\tau)_{\wp},
\end{eqnarray}
where ${\cal F}^{(p)}(\tau)_{\wp}$ is the $p$th-order contribution.
One can easily verify that ${\scal{\psi^{(0)}(\tau)}{\psi^{(0)}(\tau)}_{\wp}=1}$ [see Eq.\,\eqref{zero}], $\scal{\psi^{(0)}(\tau)}{\psi^{(p)}(\tau)}_{\wp}\!=\!b_{0}^{(p)}(\tau)_{\wp}$ [see Eq.\,\eqref{psip}], and ${{\rm Re}\,b_{0}^{(1)}(\tau)_{\wp}=0}$ [see Eq.\,\eqref{bn1}].
Thus the fidelity up to the ${P=4}$ term of Eq.\,\eqref{seriesF} is determined from
\begin{widetext}
\begin{eqnarray}\label{O3}
{\cal F}(\tau)_{\wp}\!=&&\!\biggl[1\!-\!\frac{\sum_n\bigl|b_{n}^{(1)}\bigr|^2\!\!+\!2{\rm Re}\,b_{0}^{(2)}}{T^2}
\!-\!\frac{2{\rm Re}\bigl(\sum_n b_{n}^{(1)*}b_{n}^{(2)}\!\!+\!b_{0}^{(3)}\bigr)}{T^3}
\!-\!\frac{\sum_n\bigl|b_{n}^{(2)}\bigr|^2\!\!+\!2{\rm Re}\bigl(\sum_n b_{n}^{(1)*}b_{n}^{(3)}\!\!+\!b_{0}^{(4)}\bigr)\!\!-\!\bigl(\sum_n\bigl|b_{n}^{(1)}\bigr|^2\!\!+\!2{\rm Re}b_0^{(2)}\bigr)^2}{T^4}
\nonumber\\
&&+{\cal O}\biggl(\!\frac{1}{T^5}\!\biggr)\biggr]
\biggl[1+\frac{\bigl|b_{0}^{(1)}\bigr|^2\!\!+\!2{\rm Re}\,b_{0}^{(2)}}{T^2}
\!+\!\frac{2{\rm Re}\bigl(b_{0}^{(1)*}b_{0}^{(2)}\!\!+\!b_{0}^{(3)}\bigr)}{T^3}
\!+\!\frac{\bigl|b_{0}^{(2)}\bigr|^2\!\!+\!2{\rm Re}\bigl(b_{0}^{(1)*}b_{0}^{(3)}\!\!+\!b_{0}^{(4)}\bigr)}{T^4}
+\!{\cal O}\biggl(\!\frac{1}{T^5}\!\biggr)\biggr],
\end{eqnarray}
where the first and second square brackets, respectively, correspond to the squared normalization factor and the squared sum from Eq.\,\eqref{seriesF}.
We used a shorthand notation ${b_n^{(p)}=b_n^{(p)}(\tau)_{\wp}}$ and the star for complex conjugation.
The above expression immediately yields
\begin{equation}\label{Fterms}
\begin{array}{l}
{\cal F}^{(0)}(\tau)_{\wp}=1,
\quad
{\cal F}^{(1)}(\tau)_{\wp}=0,
\quad
{\cal F}^{(2)}(\tau)_{\wp}=-\sum\limits_{n>0}\bigl|b_{n}^{(1)}(\tau)_{\wp}\bigr|^2,
\quad
{\cal F}^{(3)}(\tau)_{\wp}=-2{\rm Re}\sum\limits_{n>0}b_{n}^{(1)}(\tau)_{\wp}^*\,b_{n}^{(2)}(\tau)_{\wp},
\\
{\cal F}^{(4)}(\tau)_{\wp}=-\sum\limits_{n>0}\left[\bigl|b_n^{(2)}(\tau)_{\wp}\bigr|^2\!\!+\!2{\rm Re}\bigl\{b_n^{(1)}(\tau)_{\wp}b_n^{(3)}(\tau)_{\wp}^*\bigr\}-\bigl|b_n^{(1)}(\tau)_{\wp}\bigr|^2
\biggl(\sum\limits_{m}\bigl|b_m^{(1)}(\tau)_{\wp}\bigr|^2\!\!+\!2{\rm Re}\,b_0^{(2)}(\tau)_{\wp}\biggr)\right].
\end{array}
\end{equation}
\end{widetext}

These considerations can be directly converted to the fourth-order formula for the final ${\tau=1}$ infidelity:
\begin{equation}\label{infi2}
{\cal I}(1)_{\wp}=\frac{{\cal I}^{(2)}(1)_{\wp}}{T^2}+\frac{{\cal I}^{(3)}(1)_{\wp}}{T^3}+\frac{{\cal I}^{(4)}(1)_{\wp}}{T^4}+{\cal O}\biggl(\!\frac{1}{T^5}\!\biggr).
\end{equation}
Here we introduce the $p$th-order infidelity terms given by ${{\cal I}^{(0)}(\tau)_{\wp}=0}$ and ${{\cal I}^{(p)}(\tau)_{\wp}=-{\cal F}^{(p)}(\tau)_{\wp}}$ for ${p>0}$, see Eq.\,\eqref{Fterms}.
For very long driving times, when the leading term ${\propto T^{-2}}$ completely dominates in Eq.\,\eqref{infi2}, the final infidelity in the logarithmic form reads as
\begin{equation}\label{infilog}
\log{\cal I}(1)_{\wp}\approx-2\log T+\log\biggl[\sum_{n>0}\bigl|b_{n}^{(1)}(1)_{\wp}\bigr|^2\biggr],
\end{equation}
where the absolute term $\log{\cal I}^{(2)}(1)_{\wp}$ is identified [cf. Eq.\,\eqref{Fterms}] with the logarithm of the total first-order transition probability to all excited states.

\begin{figure}[tp]
	\includegraphics[width=\linewidth]{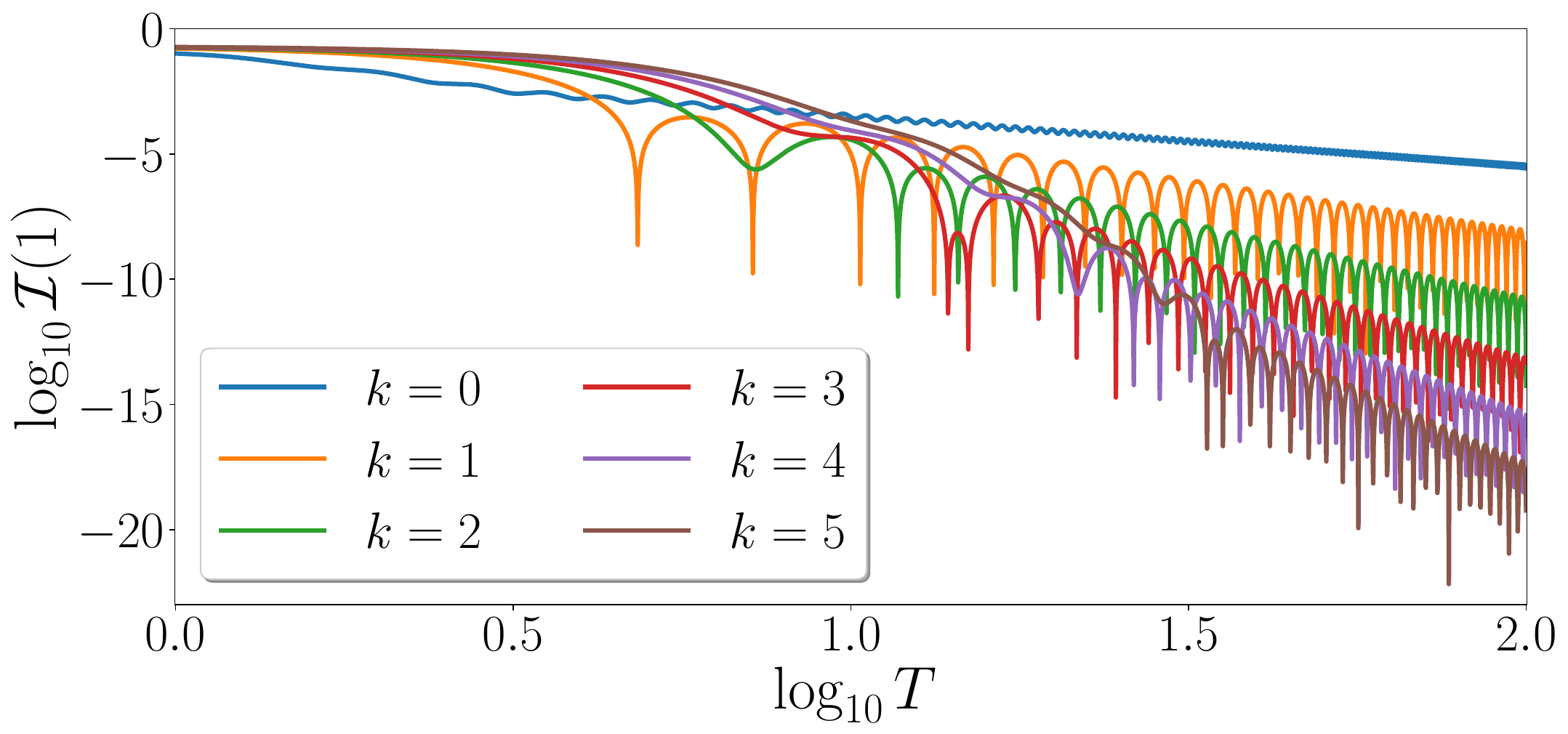}
	\caption{
(Color online) A log-log plot of the final infidelity ${{\cal I}(1)_{\wp}}$ as a function of the total driving time $T$ for various driving protocols in a two-level model explained in Sec.\,\ref{se:two}.
Hamiltonian \eqref{HLZ} is driven along line \eqref{eq:li} with ${(x_{0},z_{0})=(0.5,1)}$ and $s(\tau)$ from Table~\ref{tabul}. 
We compare linear driving protocols with ${\dot{\Lambda}^{\mu}={\rm const}\neq 0}$ (${k=0}$) and polynomial driving protocols of increasing order ${2k+1}$ ($k=1,2,\ldots$) that for the initial and final times ${\tau=0,1}$ yield $\dot{\Lambda}^{\mu}=\ddot{\Lambda}^{\mu}=\ldots=\frac{d^k}{d\tau^k}{\Lambda}^{\mu}=0$. 
The protocols with increasing $k$ yield an ordered decreasing sequence of infidelity on the rightmost side of the figure.
The slopes of the upper envelope of individual curves for ${\log T\gtrsim 1.5}$ are given by ${-2(k+1)}$.
	}
	\label{fig:poly}
\end{figure}

From formula \eqref{infilog} we conclude that for large driving times, the leading dependence of the final infidelity on time in the log-log representation is a linear decrease  with generic slope~$-2$. 
Local deviations from this behavior are caused by a residual dependence of the absolute term $\log{\cal I}^{(2)}(1)_{\wp}$ on $T$.
This conclusion is verified by numerical simulations of driven dynamics within specific models---see Fig.\,\ref{fig:poly} and the forthcoming sections.
We observe (besides the cases specified below) that in a long-time domain the $\log{\cal I}(1)_{\wp}\times\log T$ plot exhibits an upper envelope which linearly decreases with slope $-2$.
Fast oscillations with $T$ below the envelope result from the evolving phase $\varphi_{n0}(1)_{\wp}$ in $b_{n}^{(1)}(1)_{\wp}$ for ${n>0}$, which causes alternation of constructive and destructive interference of the two terms in Eq.\,\eqref{bn1} (second line).
The linearly decreasing upper envelope of the observed log-log dependence corresponds to the constructive interference of both terms, while local undershoots of the linear dependence are caused by the destructive interference.
 
The driving protocols operated in long enough times to observe the above-described linear behavior are said to be in the asymptotic time regime. 
The lower bound of the efficiency of any generic protocol in this regime is determined by the upper envelope of the $\log{\cal I}(1)_{\wp}\times\log T$ dependence.
Having reached the asymptotic time regime for several state preparation protocols based on various driving paths \eqref{path}, we can 
guarantee that the upper limits of infidelity for these protocols will not change their ordering with increasing $T$. 

Formula \eqref{infilog} is valid for a generic driving protocol, but one may ask whether it is not possible to design specific, nongeneric protocols for which one or more of low-order coefficients in Eq.\,\eqref{infi2} become zero.
Yes, it is indeed possible to design such protocols.
Looking at Eq.\,\eqref{bn1}, we notice that both coefficients ${\cal I}^{(2)}(1)_{\wp}$ and ${\cal I}^{(3)}(1)_{\wp}$ simultaneously vanish if
\begin{equation}\label{speed0}
\dot{\Lambda}^{\mu}(0)=\dot{\Lambda}^{\mu}(1)=0,\quad\mu=1,2,\ldots,D,
\end{equation}
that is, if the instantaneous speed is zero at the beginning as well as at the end of the driving path.
In this case we have ${b_n^{(1)}(1)_{\wp}=0}$ for ${n>0}$, which according to Eq.\,\eqref{Fterms} yields ${{\cal F}^{(2)}(1)_{\wp}={\cal F}^{(3)}(1)_{\wp}=0}$.
For driving protocols satisfying condition \eqref{speed0} we can write
\begin{equation}\label{infilog4}
\log{\cal I}(1)_{\wp}\approx-4\log T+\log\biggl[\sum_{n>0}\bigl|b_{n}^{(2)}(1)_{\wp}\bigr|^2\biggr],
\end{equation}
so the infidelity logarithm in the asymptotic time regime decreases with $\log T$ linearly with slope~$-4$ instead of~$-2$, and the absolute term is the logarithm of the total second-order transition probability to all excited states.

For the paths satisfying condition \eqref{speed0} one can derive the following expression for ${n>0}$ coefficients in Eq.\,\eqref{infilog4}:
\begin{equation}\label{eq:bn2_N4}
b_n^{(2)}(1)=-e^{i\varphi_{n0}(1)_{\wp}}\frac{\frac{d}{d\tau}\frac{M_{n0}(\tau)}{\Delta_{n0}(\tau)}\bigr|_{\tau=1}}{\Delta_{n0}(1)}+\frac{\frac{d}{d\tau}\frac{M_{n0}(\tau)}{\Delta_{n0}(\tau)}\bigr|_{\tau=0}}{\Delta_{n0}(0)}
\end{equation}
(cf.\,Ref.\,\cite{Orti08}).
These coefficients become zero for driving protocols satisfying an additional condition
\begin{equation}\label{accel0}
\ddot{\Lambda}^{\mu}(0)=\ddot{\Lambda}^{\mu}(1)=0,\quad\mu=1,2,\ldots,D,
\end{equation}
which sets vanishing initial and final accelerations. 
Hence for driving protocols satisfying simultaneously Eqs.\,\eqref{speed0} and \eqref{accel0}, the dependence of the final infidelity on $T$ in the asymptotic time regime is pushed to even higher terms than the fourth-order one.
Although we have not derived explicit analytic formulas for higher-order infidelity terms, we have checked numerically that this game can be played repeatedly: for higher vanishing derivatives of $\Lambda^{\mu}$ at ${\tau=0}$ and ${\tau=1}$ we gain decreasing negative slopes of the asymptotic log-log dependence of the final infidelity on the driving time.
This is illustrated in Fig.\,\ref{fig:poly}, which compares the final infidelity in a toy model from Sec.\,\ref{se:two} for driving protocols satisfying ${\frac{d^i}{d\tau^i}\bLambda(0)=\frac{d^i}{d\tau^i}\bLambda(1)=0}$ with ${i=1,2,\ldots,k}$.
We observe that the slope of the $\log{\cal I}(1)_{\wp}\times\log T$ plot in the asymptotic time regime (or more precisely, the slope of the upper envelope of an oscillatory dependence) takes values $-4,-6,-8\ldots$ for protocols with $k=1,2,3,\ldots$, respectively.

We conclude that the drivings with vanishing initial and final derivatives of $\Lambda^{\mu}(\tau)$ lead to a reduction of the maximum final infidelity for very large driving times.
The more derivatives vanish, the better result can be reached. 
This potentially represents a very useful technique for designing optimal state preparation protocols.
However, it needs to be stressed that for $T$ before or at the beginning of the asymptotic time regime the protocols with vanishing derivatives may yield worse results than some other protocols.
This is seen already in Fig.\,\ref{fig:poly} and will be further illustrated below.

\section{Driving protocols}
\label{sec:drivings}

\begin{figure}[tp]
	\includegraphics[width=0.85\linewidth]{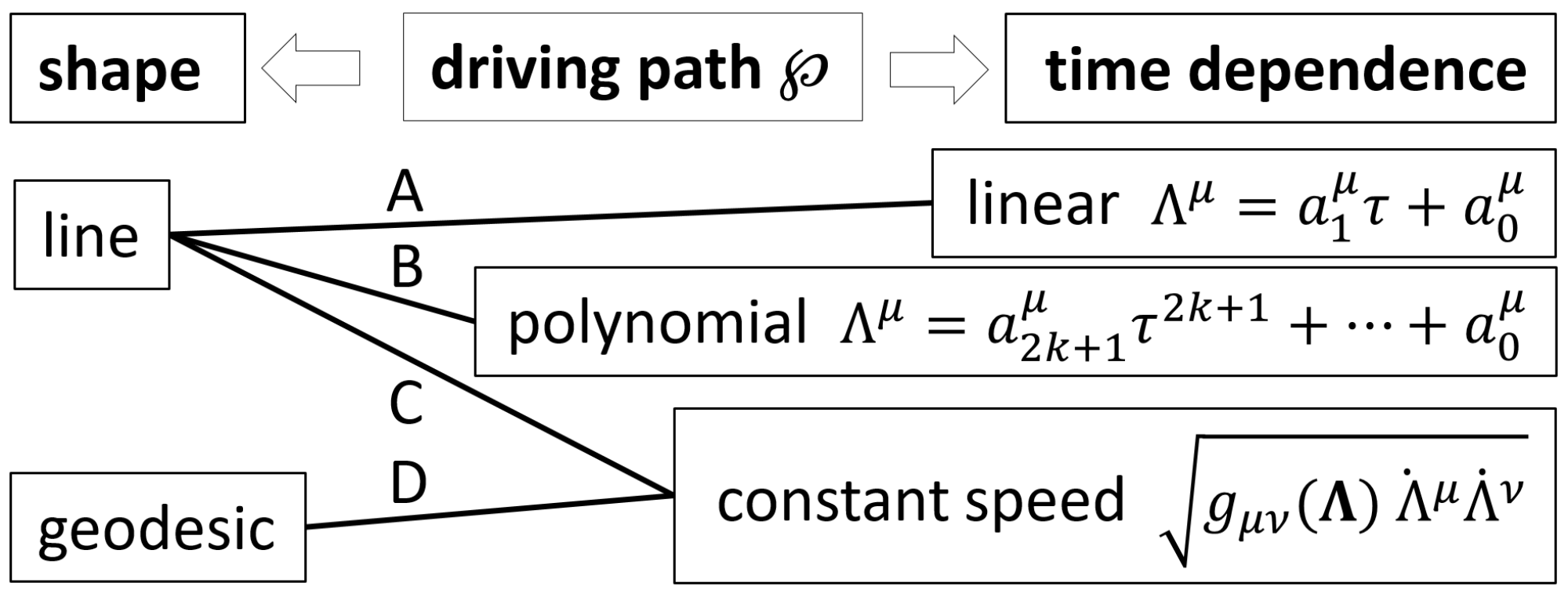}
	\caption{
Driving protocols explained in Sec.\,\ref{sec:drivings}. Path \eqref{path} connecting points $\bLambi$ and $\bLambf$ in the parameter space is characterized by its shape  (specified on the left) and by its time dependence or speed (specified on the right).
The coefficients $a_i^{\mu}$ in the linear and polynomial drivings follow from formula~\eqref{lindri} and Table~\ref{tabul}. 
Links denoted by letters~A, B, C, D define four combinations employed below.
	}
	\label{fig:dri}
\end{figure}

For any Hamiltonian $\hat{H}(\bLambda)$, there exist an infinite number of driving protocols, i.e., specific time dependencies $\Lambda^{\mu}(t)$, that take us from a selected initial point $\bLambi$ to a desired final point $\bLambf$ in a given total time $T$.
The question is which of these protocols yields larger fidelity with respect to the final ground state $\ket{E_0(\bLambf)}$.  
In this section, we describe the driving protocols tested in our work.
A sketch of these protocols is presented in Fig.\,\ref{fig:dri}.
Each protocol is characterized by a shape of the corresponding curve in the parameter space and by a time dependence of the motion along this curve. 
We first introduce the linear and polynomial driving protocols and then explain the constant-speed and geodesic protocols based on the geometric structure of the parameter space.

\subsection{Linear and polynomial drivings}
\label{se:lin}

The simplest way of getting from $\bLambi$ to $\bLambf$ is to go along a straight line in the parameter space.
The line is determined by
\begin{equation}\label{lindri}
\Lambda^{\mu}(s)=(\Lambf^{\mu}-\Lambi^{\mu})\,s+\Lambi^{\mu},\quad s\in[0,1],
\end{equation}
where $s$ is a parameter specifying a fraction of the line already passed.
This parameter is supposed to depend on the scaled time $\tau$ so that $s(\tau)$ monotonously increases from~0 to~1 as $\tau$ runs from~0 to~1.
In the simplest case, hereafter called the linear driving, we set ${s(\tau)=\tau}$.
This yields constant derivatives $\dot{\Lambda}^{\mu}(\tau)={\Lambf^{\mu}-\Lambi^{\mu}}$ and $\frac{d^i}{d\tau^i}\Lambda^{\mu}(\tau)=0$ for ${i>1}$.

In order to implement the driving protocols with vanishing derivatives at the initial and final times (see Sec.\,\ref{sub:apliAPT}), we apply a~straightforward generalization of the linear driving called a polynomial driving.
In this case, the dependence of $s(\tau)$ is given by a polynomial of an odd order ${2k+1}$,
\begin{equation}\label{poldri}
 	s(\tau) = s_{2k+1}\tau^{2k+1}+s_{2k}\tau^{2k}+\ldots+s_{1}\tau,
 \end{equation}
where $\{s_n\}_{n=1}^{2k+1}$ are coefficients that need to be fixed with respect to the required cancellation of some of the ${\tau=0,1}$ derivatives.
Vanishing of the derivatives up to the $k$th term sets $2k$ constraints, an additional constraint follows from the condition ${s(1)=1}$, while the last condition ${s(0)=0}$ is guaranteed by the missing absolute term~$s_0$. 
The forms of the polynomial \eqref{poldri} for values of $k$ up to 5 are given in Table \ref{tabul}.
We note that the linear driving is apparently a special case of the polynomial driving with ${k=0}$. 

\begin{table}
\begin{ruledtabular}
\begin{tabular}{ll}
$k$ & $s(\tau)$ 
\\
\hline
0 &  $\tau$ \qquad\qquad\qquad\qquad\qquad\qquad\quad linear \\
1 &  $2\bigl(-\tau^3+\frac{3}{2}\tau^2\bigr)$ \qquad\qquad\qquad\quad\,\  polynomial \\
2 &  $6\bigl(\tau^5-\frac{5}{2}\tau^4+\frac{5}{3}\tau^3\bigr)$ \\
3 &  $20\bigl(-\tau^7+\frac{7}{2}\tau^6-\frac{21}{5}\tau^5+\frac{7}{4}\tau^4\bigr)$ \\
4 &  $70\bigl(\tau^9-\frac{9}{2}\tau^8+\frac{54}{7}\tau^7-6\tau^6+\frac{9}{5}\tau^5\bigr)$\\
5 &  $252\bigl(-\tau^{11}+\frac{11}{2}\tau^{10}-\frac{110}{9}\tau^9+\frac{55}{4}\tau^8-\frac{55}{7}\tau^7+\frac{11}{6}\tau^6\bigr)$\\
\end{tabular}
\end{ruledtabular}
\caption{Polynomials \eqref{poldri} satisfying ${s(1)=1}$ and yielding zero derivatives $\frac{d^i}{d\tau^i}\Lambda^{\mu}(\tau)$ at ${\tau=0}$ and 1 for $i=1,2,\ldots,k$.}
\label{tabul}
\end{table}

\subsection{Drivings based on the geometric structure}
\label{se:ge}

Following the approach initiated by Provost and Val\-lee~\cite{Prov80} and extended by Berry and others~\cite{Berr84,Wilc88,Berr88,Anan90,Kuma12,Kolo17}, one can equip the parameter space of Hamiltonian $\hat{H}(\bLambda)$ with a geometric structure invoking the formalism of curved spaces.
Since recent literature presents several attempts to apply this formalism to the design of optimal state preparation protocols \cite{Buko19,Kolo17,Tomk16}, we also include the geometric approach to the present analysis.

Consider an infinitely small shift in the parameter space from $\bLambda$ to $\bLambda+d \bLambda$, where
$d\bLambda\equiv{(d\Lambda^1,d\Lambda^2,\ldots,d\Lambda^D)}$ has $D$ infinitesimal components $d\Lambda^{\mu}$.
The geometric approach starts from the definition of an element of distance $d\ell$ associated with this shift.
It does not measure just a length covered in the parameter space, but reflects the induced change of the system properties, particularly the modification of individual Hamiltonian eigenstates. 
Considering the $n$th eigenstate, we associate with each parameter point $\bLambda$ a set $\{e^{i\gamma(\bLambda)}\ket{E_n(\bLambda)}\}$ of vectors differing by phase factors with arbitrary ${\gamma(\bLambda)\in[0,2\pi)}$.
This defines a fibered manifold, hereafter called the $n$th-state manifold. 
As we intend to maximize the fidelity of the ground state, we focus on the ground-state manifold. 
The squared element of distance $d\ell^2$ on the ground-state manifold is given by
\begin{equation}
{d\ell}^{2} = 1-\left|\scal{E_0(\bLambda)}{E_0(\bLambda\!+\!d \bLambda)}\right|^2  
=g_{\mu\nu}(\bLambda)\ d\Lambda^{\mu} d\Lambda^{\nu},
\label{eq:dl}
\end{equation}
which is apparently independent of any local choice of phases~$\gamma(\bLambda)$.
The first expression identifies the squared distance with the ground-state infidelity generated by a~sudden jump of parameters  from $\bLambda$ to ${\bLambda+d \bLambda}$ (an infinitesimal quench). 
The second expression introduces the metric tensor $g_{\mu\nu}(\bLambda)$ \cite{Prov80}.
It can be determined from
\begin{eqnarray}
g_{\mu\nu}\!&=&\!{\rm Re}\left[
\scal{\tfrac{\partial}{\partial\Lambda^{\mu}}E_0}{\tfrac{\partial}{\partial\Lambda^{\nu}}E_0}\!-\!\scal{\tfrac{\partial}{\partial\Lambda^{\mu}}E_0}{E_0}\scal{E_0}{\tfrac{\partial}{\partial\Lambda^{\nu}}E_0}\right]
\nonumber\\
&=&{\rm Re}\sum_{n>0} \frac{\matr{E_0}{\frac{\partial}{\partial\Lambda^{\mu}}\hat{H}}{E_n} \matr{E_n}{\frac{\partial}{\partial\Lambda^{\nu}}\hat{H}}{E_0}}{\Delta_{n0}^2},
\label{metr}
\end{eqnarray}
where, for simplicity, we suppressed marking of the dependencies on $\bLambda$ (as in some formulas below).

The metric tensor $g_{\mu\nu}$ is naturally symmetric under the exchange of indices $\mu$ and $\nu$ since any antisymmetric part would not contribute to the expression \eqref{eq:dl}. 
It can be supplemented by an antisymmetric component, proportional to the imaginary part of the expressions in Eq.\,\eqref{metr}. 
This so-called curvature tensor determines geometric phases acquired in adiabatic drivings along closed paths~\cite{Berr84}. 

The norm of the metric tensor is large in those parameter regions where the energy gaps $\Delta_{n0}$ are small and/or where the matrix elements $\matr{E_0}{\frac{\partial}{\partial\Lambda^{\mu}}\hat{H}}{E_n}$ are large.
These are the most problematic regions for drivings whose aim is to minimize the ground-state infidelity.
As we see from Eq.\,\eqref{eq:dl}, the squared distance element $d\ell^2$ measures the ground-state infidelity caused by an infinitesimal parameter quench, so it roughly reflects the difficulty to transfer the state over the corresponding interval in the parameter space.
Limitations of this statement will be discussed below.

Having defined the metric on the ground-state manifold, we can measure the length $\ell$ of an arbitrary stretch of any curve ${\cal C}$ in the parameter space (${\cal C}$ defines only the shape of the path $\wp$ and not the time dependence of driving along it). 
Let $\bLambda(s)$, $s\in[0,1]$ be a parametrization of such a curve.
Then 
\begin{equation}\label{length}
\ell(s)=\int\limits_0^s \sqrt{g_{\mu\nu}\bigl(\bLambda(s')\bigr)\frac{d\Lambda^{\mu}(s')}{ds'}\frac{d\Lambda^{\nu}(s')}{ds'}}\ ds'
\end{equation}
measures the length of the stretch of ${\cal C}$ from the start ${s=0}$ to the point corresponding to a given $s$.
Prescribing to the curve parameter an arbitrary time dependence~$s(\tau)$, we can determine an instantaneous speed at any moment of the driving along ${\cal C}$, 
\begin{equation}\label{spee}
\dot{\ell}(\tau)=\sqrt{g_{\mu\nu}(\tau)\dot{\Lambda}^{\mu}(\tau)\dot{\Lambda}^{\nu}(\tau)},
\end{equation}
where ${g_{\mu\nu}(\tau)=g_{\mu\nu}(\bLambda(\tau))}$.

From definition \eqref{eq:dl} we can expect that the speed on the manifold \eqref{spee} reflects an instantaneous transition rate from the ground state to all excited states.
Therefore, to minimize the losses of fidelity, it may be useful to avoid any maxima of the speed and perform the driving in so that $\dot{\ell}(\tau)$ remains constant, fixed by a given total duration $T$.
This idea can be implemented for the above-discussed driving protocols performed along a line in the parameter space (Sec.\,\ref{se:lin}).
Adopting parametrization~\eqref{lindri} and applying the condition ${\dot{\ell}(\tau)={\rm const}}$, we determine the dependence~$s(\tau)$ from an implicit equation
\begin{equation}\label{vconst}
\underbrace{\int\limits_{0}^{s} ds'\sqrt{g_{\mu\nu}\bigl(\bLambda(s')\bigr)(\Lambf^{\mu}\!-\!\Lambi^{\mu})(\Lambf^{\nu}\!-\!\Lambi^{\nu})}}_{A(s)}=A(1)\,\tau.
\end{equation}
This type of driving complements the previously discussed linear and polynomial drivings from Table~\ref{tabul}.
So, in connection to the paths whose shape is a line, we have three types of driving protocols (see Fig.\,\ref{fig:dri}): linear (type A), polynomial of various orders (type B), and the constant-speed one (type C).

The above-mentioned interpretation of $d\ell^2$ as an infidelity generated by an infinitesimal quench gives rise to a question whether the fidelity of the state preparation procedure in time $T$ can be further improved (on top of the possible  improvement due to the constant speed condition) by reducing the total length ${\cal L}=\ell(1)$ of the corresponding curve.
Is the geodesic, i.e., the curve of minimal length among the curves connecting points $\bLambi$ and $\bLambf$, always an optimal trajectory for the ${\dot{\ell}(\tau)={\rm const}}$ driving?
The negative answer to this question based on our numerical simulations will be manifested below.
Nevertheless, some skepticism to such a direct link between geometry and dynamics follows already from general arguments, namely from the fact that full quantum evolution by time-dependent Hamiltonians unavoidably involves coherence effects which are not properly reflected in the quench picture behind Eq.\,\eqref{eq:dl}.
This will be further commented in Secs.\,\ref{se:re1} and \ref{se:re2}.

The geodesic can be obtained by variation of Eq.\,\eqref{length} with ${s=1}$. 
This yields the differential equation
\begin{eqnarray}
	\frac{d^2\Lambda^{\mu}}{ds^2} + &\Gamma^{\mu}_{\ \nu\rho}& \frac{d\Lambda^{\nu}}{ds}\frac{d\Lambda^{\rho}}{ds} = 0,
	\label{geoeq}
	\\
	&\Gamma^{\mu}_{\ \nu\rho}& = \frac{1}{2} g^{\mu\xi}\biggl(\frac{\partial g_{\xi\nu}}{\partial\Lambda^{\rho}} + \frac{\partial g_{\xi\rho} }{\partial\Lambda^{\nu}}- \frac{\partial g_{\nu\rho}}{\partial \Lambda^{\xi}}\biggr),
	\nonumber
\end{eqnarray}
where $\Gamma^{\mu}_{\ \nu\rho}$  are the Christoffel symbols of the second kind and $g^{\mu\xi}$ are components of the inverse metric tensor~\cite{Pate00}.
It can be shown that the solution to Eq.\,\eqref{geoeq} automatically satisfies the condition ${\frac{d}{ds}\ell={\rm const}}$.
This means that using the simplest time dependence of the curve parameter~$s$, namely ${s(\tau)=\tau}$, we select the paths that follow the geodesic curve and in addition keep a constant speed on the manifold, ${\dot{\ell}(\tau)={\rm const}}$.
These represent the fourth type of the drivings employed below (type D in Fig.\,\ref{fig:dri}).

\section{Driving in a one-qubit system}
\label{se:two}

So far, we have discussed the problem of driving independently of a specific system where it is realized.
In this and the following sections we apply the general ideas to concrete systems, in which we can identify some precursors of ground-state QPTs.
We start with an elementary system that hints at such behavior, namely a single-qubit model with an avoided crossing of energy levels.

\subsection{The model}
\label{Twolev}

Inspired by Ref.\,\cite{Tomk16}, we use the well-known Landau-Zener model~\cite{Landau} defined in a two-dimensional Hilbert space ${{\cal H}=\mathbb{C}^2}$ spanned by vectors $\left(\begin{smallmatrix}1\\0\end{smallmatrix}\right)$ and $\left(\begin{smallmatrix}0\\1\end{smallmatrix}\right)$. 
These can be interpreted as the basis states $\ket{0}$ and $\ket{1}$ of a~qubit.
The Hamiltonian is considered real, given by 
\begin{equation}\label{HLZ}
	\hat{H} =  x\,\hat{\sigma}_x + z\,\hat{\sigma}_z=
	\begin{pmatrix}
		z & x \\
		x & -z
	\end{pmatrix}=
	r\begin{pmatrix}
		\sin\alpha & \cos\alpha \\
		\cos\alpha & -\sin\alpha
	\end{pmatrix}
	,
\end{equation}
where $\hat{\sigma}_x$ and $\hat{\sigma}_z$ stand for Pauli matrices, and $x,z\in(-\infty,+\infty)$  are parameters characterizing an external field that will be later subject to a prescribed time dependence $\bLambda(\tau)=\bigl(x(\tau),z(\tau)\bigr)$.

The parametrization of Hamiltonian \eqref{HLZ} by $(x,z)$ is apparently redundant since the expression in terms of polar coordinates ${r\in[0,\infty)}$ and ${\alpha\in[0,2\pi)}$ reveals that~$r$ represents merely a scaling factor.
Moreover, any Hamiltonians with the same $r$ and different angles $\alpha$ and $\alpha'$ are equivalent up to a unitary transformation $\hat{U}=e^{-i\hat{\sigma}_y(\alpha-\alpha')/2}$.
The Hamiltonian eigenvalues and eigenvectors read as 
\begin{eqnarray}
E_0=-r,&&\quad \ket{E_0}=\frac{1}{\sqrt{2}}\left(\!\!\begin{array}{r}-\eta_1(\alpha)\sqrt{1\!-\!\sin\alpha}\\\eta_2(\alpha)\sqrt{1\!+\!\sin\alpha}\end{array}\right),
\label{E1}\\
E_1=+r,&&\quad \ket{E_1}=\frac{1}{\sqrt{2}}\left(\begin{array}{r}\eta_2(\alpha)\sqrt{1\!+\!\sin\alpha}\\\eta_1(\alpha)\sqrt{1\!-\!\sin\alpha}\end{array}\right),
\label{E2}
\end{eqnarray}
where ${\eta_1(\alpha)={\rm sgn}(\frac{\pi}{2}\!-\!\alpha)}$ and ${\eta_2(\alpha)={\rm sgn}(\frac{3\pi}{2}\!-\!\alpha)}$.

The parameter plane of Hamiltonian \eqref{HLZ} contains a single diabolic point $(x,z)=(0,0)$ where both eigenvalues coincide.
Selecting as the driving path in the plane ${x\times z}$ any line passing near the origin, we let the system evolve through an avoided crossing of both energy levels, the spacing ${\Delta_{10}=2r}$ being minimal when the radius $r$ takes the smallest value.
This can be considered as a toy model for simulating precursors of quantum phase transitions in finite-size systems.

The metric tensor expressed in the $(x,y)$ and $(r,\alpha)$ coordinates has the form 
\begin{equation}
\hspace{-1.5mm}
g_{\mu\nu}\!\equiv\!\left\{\!\!
\begin{array}{l}
\left(\begin{array}{cc}g_{xx}&g_{xz}\\g_{zx}&g_{zz}\end{array}\right)\!={\displaystyle \frac{1}{4(x^2\!+\!z^2)^2}}\left(\begin{array}{cc}z^2&-xz\\-xz&x^2\end{array}\right),
\\
\left(\begin{array}{cc}g_{rr}&g_{r\alpha}\\g_{\alpha r}&g_{\alpha\alpha}\end{array}\right)\!={\displaystyle\frac{1}{4}}\left(\begin{array}{cc}0&0\\ 0&1\end{array}\right).
\end{array}\right.
\label{gLZ}
\end{equation} 
Identities ${g_{rr}=g_{r\alpha}=g_{\alpha r}=0}$, following directly from the independence of eigenvectors on $r$, imply a vanishing eigenvalue of $g_{\mu\nu}$.
This means that the model has a singular metric structure.
Indeed, the length of an arbitrary curve $\bigl(r(s),\alpha(s)\bigr)$ depends only on the angular part $\alpha(s)$, so for any pair of points in the parameter plane there exist infinite number of connecting curves of the same length.
Also the geodesic between these points is undetermined, including for instance linear and arc paths, as well as multitudes of paths showing arbitrary wiggles in the radial direction.

\subsection{Results and discussion}
\label{se:re1}

\begin{figure}[tp]
	\includegraphics[width=\linewidth]{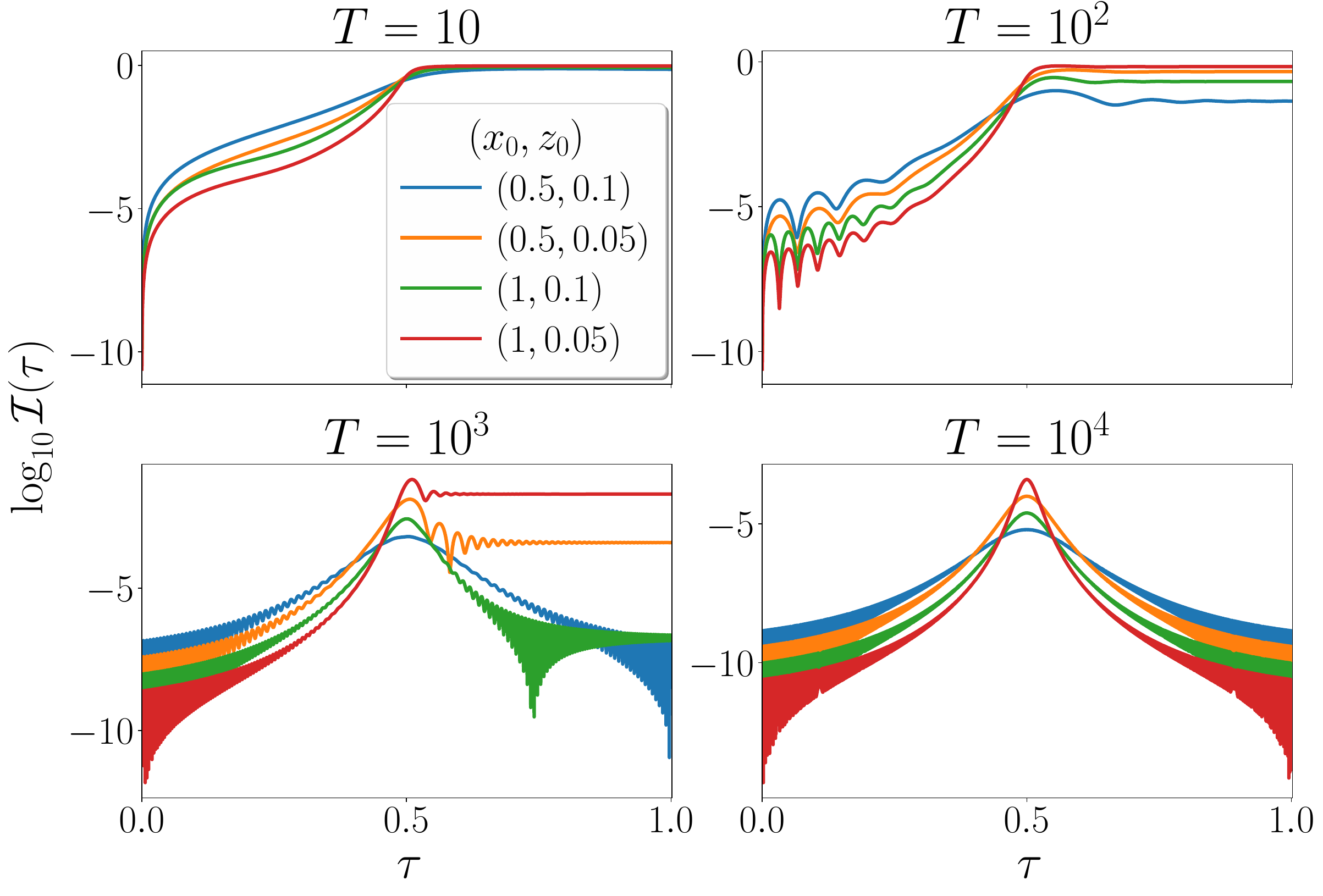}
	\caption{
(Color online) Instantaneous infidelity \eqref{fidel} at time $\tau$ in the two-level model for linear driving protocols~A along line \eqref{eq:li} with ${s=\tau}$ for four values of $(x_{0},z_{0})$ (the ordering of curves in the left part of each panel corresponds to the ordering of the legend) and four total driving times $T$ (various panels).
	}
	\label{fig:infti1}
\end{figure}

Despite its simplicity, the two-level model demonstrates majority of intricacies involved in the problem of driven quantum dynamics.
In the Hilbert space of dimension two it is possible to find various analytical and approximative solutions to the nonstationary Schr{\"o}dinger equation, see, e.g., Refs.\,\cite{Vita96,Vita99,Yan10}. 
Here we use mostly numerical (hence exact) calculations, as well as explicit evaluation of some APT expressions. 
We nevertheless start with some differential equations governing the evolution of fidelity ${\cal F}(\tau)_{\wp}$ and infidelity ${{\cal I}(\tau)_{\wp}=1-{\cal F}(\tau)_{\wp}}$ from Eq.\,\eqref{fidel}.
For ${{\cal F}(\tau)_{\wp},{\cal I}(\tau)_{\wp}\neq\{0,1\}}$ we obtain 
\begin{eqnarray}\label{diFI}
\dot{{\cal F}}(\tau)_{\wp}&&=-\dot{{\cal I}}(\tau)_{\wp}=
\\
&&2\bigl|M_{10}(\tau)\bigr|\sqrt{{\cal F}(\tau)_{\wp}{\cal I}(\tau)_{\wp}}\cos{\bigl(\phi_{10}(\tau)_{\wp}\!\!-\!\chi(\tau)\bigr)},
\nonumber
\end{eqnarray}
where ${M_{10}(\tau)=|M_{10}(\tau)|e^{i\chi(\tau)}}$ is defined in Eq.\,\eqref{Mnm} and the angle ${\phi_{10}(\tau)_{\wp}={\rm arg}\bigl(a_1(\tau)_{\wp}a_0(\tau)_{\wp}^*\bigr)}$ is a relative phase between complex coefficients 
in the expansion 
\begin{equation}\label{a0a1}
\ket{\psi(\tau)}_{\wp}=a_0(\tau)_{\wp}\ket{E_0(\tau)}+a_1(\tau)_{\wp}\ket{E_1(\tau)}.
\end{equation}
Note that the exact phase $\phi_{10}(\tau)_{\wp}$ should not be confused with the adiabatic phase $\varphi_{10}(\tau)_{\wp}$ from Eq.\,\eqref{phadi}.
The exact phase in our two-level system follows the equation
\begin{eqnarray}
\dot{\phi}_{10}(\tau)_{\wp}=-\Delta(\tau)T-{\rm Im}M_{11}(\tau)+{\rm Im}M_{00}(\tau)
\qquad\quad
\nonumber\\
+\bigl|M_{10}(\tau)\bigr|\frac{{\cal F}(\tau)_{\wp}\!-\!{\cal I}(\tau)_{\wp}}{\sqrt{{\cal F}(\tau)_{\wp}{\cal I}(\tau)_{\wp}}}\sin{\bigl(\phi_{10}(\tau)_{\wp}\!\!-\!\chi(\tau)\bigr)},\quad
\label{phi10}
\end{eqnarray}
where ${\Delta(\tau)\equiv\Delta_{10}(\tau)}$.
We point out that for the real Hamiltonian \eqref{HLZ} the equations \eqref{diFI} and \eqref{phi10} have a simpler form with ${M_{00}(\tau)=M_{11}(\tau)=0}$ and ${\chi(\tau)=0}$ or $\pi$. 
The nonlinearity of these equations hints at nontrivial solutions, which will be numerically confirmed below.

We also mention that in the two dimensional case there exists a simple relation between the (in)fidelity and the energy variance ${{\cal V}(\tau)_{\wp}\equiv\overline{E^2}(\tau)_{\wp}-\overline{E}(\tau)_{\wp}^2}$, where ${\overline{E^n}(\tau)_{\wp}=\matr{\psi(\tau)}{\hat{H}(\tau)^n}{\psi(\tau)}_{\wp}}$ stands for the $n$th statistical moment of energy at time $\tau$.
In particular, we have
\begin{equation}\label{variance}
{\cal V}(\tau)_{\wp}={\cal F}(\tau)_{\wp}\,{\cal I}(\tau)_{\wp}\,\Delta(\tau)^2,
\end{equation}
which indicates that for small values of infidelity (below $0.5$) the energy variance can be used as an alternative measure of driving-induced excitation.

Now let us turn to the numerical results.
The initial and final parameter points ${\bLambi\equiv(x_{\rm I},z_{\rm I})}$ and $\bLambf\equiv{(x_{\rm F},z_{\rm F})}$ in all driving protocols employed here are selected symmetrically, lying on a circle with the same radius in the ${x\times z}$ plane.
In particular we will have ${x_{\rm I}=-x_{\rm F}<0}$ and ${z_{\rm I}=z_{\rm F}>0}$, so both initial and final points are determined by $(x_{\rm F},z_{\rm F})\equiv(x_{0},z_{0})$. 
If these points are connected by a line 
\begin{equation}
\bigl(x(s),z(s)\bigr)=\bigl(\,2x_{0}\,s\!-\!x_{0},\,z_{0}\,\bigr),
\quad s\in[0,1],
\label{eq:li}
\end{equation}
the energies from Eqs.\,\eqref{E1}--\eqref{E2} exhibit a symmetric dependence ${E_0(s)=-\sqrt{x(s)^2+z^2_0}=-E_1(s)}$ with an avoided level crossing centered precisely at the halfway ${s=0.5}$ where the line gets closest to the diabolic point ${(x,z)=(0,0)}$.
It should be stressed that driving along paths obtained by arbitrary rotations of line \eqref{eq:li} around the origin would yield identical results.

Figure~\ref{fig:infti1} shows the evolution of instantaneous infidelity ${\cal I}(\tau)_{\wp}$ along trajectory \eqref{eq:li} for a linear driving with ${s=\tau}$ (case~A in Fig.\,\ref{fig:dri}) for various choices of $(x_{0},z_{0})$ and for various total driving times~$T$.
To emphasize that now we deal with linear driving protocols, we substitute $\wp={\rm A}$.
The maximum of infidelity at ${\tau=0.5}$, which appears in a more or less sharp form for most of the curves in Fig.\,\ref{fig:infti1}, corresponds to driving through the region around the minimum energy gap.
We stress that nonmonotonous dependences of infidelity on~$\tau$ follow from evolving relative phase $\phi_{10}(\tau)_{\wp}$ in Eq.\,\eqref{diFI} and hence directly reflect the coherence of quantum evolution \eqref{a0a1} of our isolated system.
Repeated passage of the phase  through the interval $[0,2\pi)$  leads to infidelity oscillations.  
From Eqs.\,\eqref{diFI},\,\eqref{phi10} and~\eqref{Mnm} we infer that the impact of phase variations on the infidelity evolution is strong if the energy gap $\Delta(\tau)$ is small, and also that a fast change of the phase is most likely if $\Delta(\tau)$ or ${\cal I}(\tau)_{\wp}$ is small.
These notes explain some features of the observed dependencies. 

\begin{figure}[tp]
	\includegraphics[width=\linewidth]{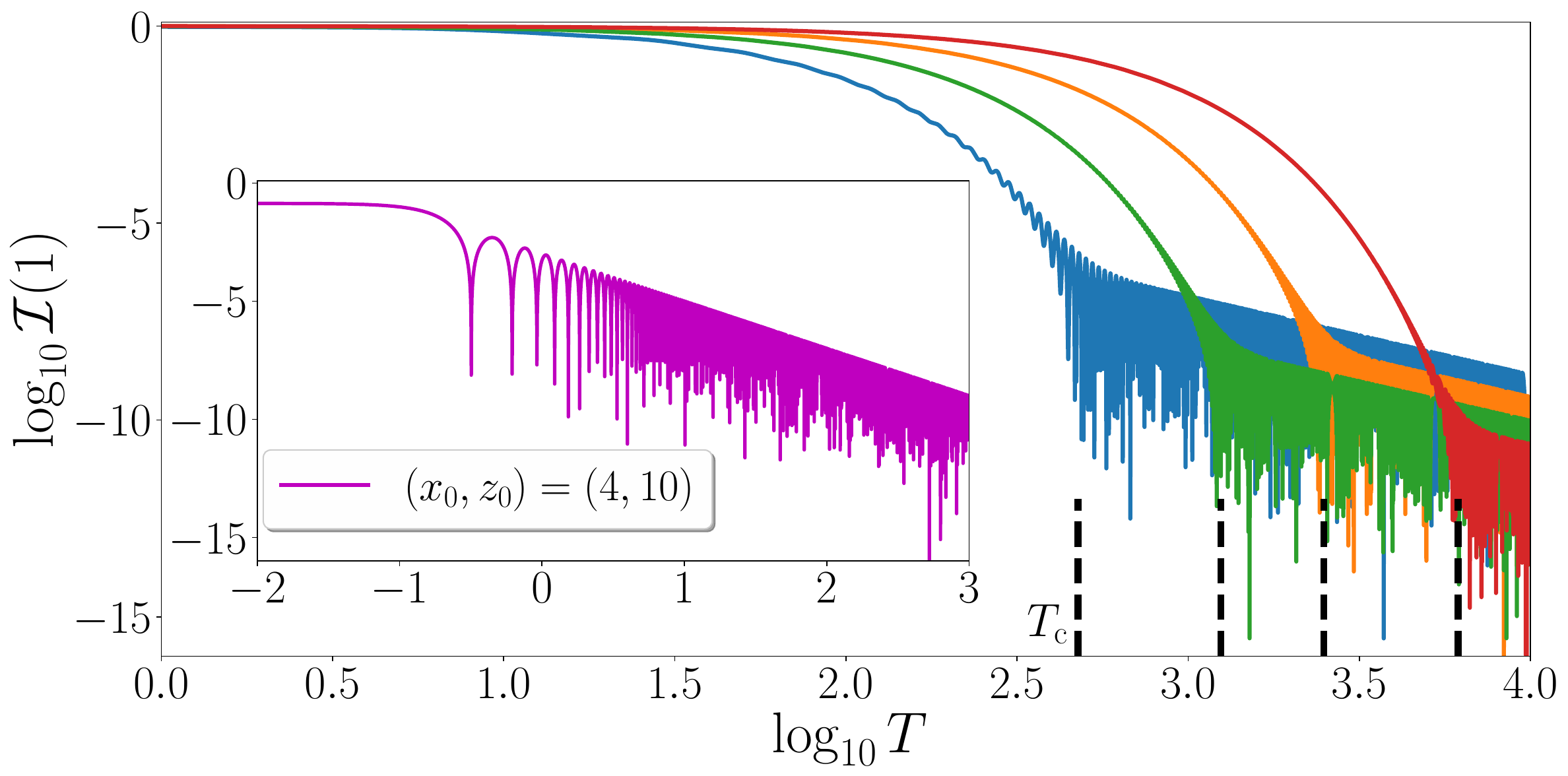}
	\caption{
(Color online) The final infidelity in the two-level model for linear driving protocols~A with the same parameters as in Fig.\,\ref{fig:infti1} as a function of total driving time~$T$ (the ordering of curves in the rightmost side of the main panel corresponds to the ordering of the legend in Fig.\,\ref{fig:infti1}).
The transition between the Landau-Zener and the asymptotic time regimes is indicated by the vertical dashed lines obtained from formula~\eqref{Lambert}.
The inset shows an example of driving that violates condition~\eqref{limra} and therefore shows only a gradual transition to the asymptotic regime.   
	}
	\label{fig:infLZ}
\end{figure}

Behind the avoided crossing at ${\tau=0.5}$, the infidelity oscillations (if any) in Fig.\,\ref{fig:infti1} are damped and for not too large values of~$T$, depending on parameters, the infidelity saturates at a final value ${\cal I}(1)_{\rm A}$ (see the curves for ${T=10},100$ and some curves for ${T=1000}$).
In these cases, the saturation value of infidelity is very well approximated by the Landau-Zener formula \cite{Landau}
\begin{equation}
\ln{\cal I}(1)_{\rm A}=
-\frac{\pi z_{0}^2}{2x_{0}}\,T,
\label{eq:LZ}
\end{equation}
where the parameters $(x_0,z_0)$ can be expressed from general \uvo{observables} describing the avoided crossing, namely the minimal energy gap, ${{\rm min\,}\Delta(\tau)=2z_0}$, and the asymptotic gap derivative, ${\lim_{\tau\to\pm\infty}\dot{\Delta}(\tau)=4x_0}$.

The exponential decrease of the infidelity with $T$ in Eq.\,\eqref{eq:LZ} looks contradictory to the results of Sec.\,\ref{sub:apliAPT}, where we showed that in the asymptotic regime the infidelity exhibits an algebraic decrease with~$T$
(see Fig.\,\ref{fig:poly}).
The resolution of this inconsistency is illustrated by Fig.\,\ref{fig:infLZ}.
It shows that the infidelity actually manifests both regimes: the Landau-Zener regime for smaller values of $T$ and the asymptotic regime for larger values of $T$.
The value $T_{\rm c}$, at which the exponential and algebraic dependencies merge, represents a crossover time that marks the beginning of the asymptotic regime.
Examples of driving in this regime are the nonsaturating curves in Fig.\,\ref{fig:infti1}. 

In the two-level model, the crossover time $T_{\rm c}$ can be easily estimated.
Indeed, the sum in Eq.\,\eqref{infilog} contains only a single term, which can be evaluated with a~help of Eqs.\,\eqref{bn1} and \eqref{HLZ}--\eqref{E2}. 
This yields 
\begin{equation}
\bigl|b_1^{(1)}(1)_{\rm A}\bigr|^2=\frac{x_{0}^2z_{0}^2}{(x_{0}^2+z_{0}^2)^3}\sin^2\frac{\varphi_{10}(1)_{\rm A}}{2},
\label{critim}
\end{equation}
where we can replace the squared sine by its average~$\frac{1}{2}$, anticipating fast oscillations of quantity~\eqref{critim} in the time domain of interest.
Equating then the Landau-Zener formula \eqref{eq:LZ} with the asymptotic formula \eqref{infilog}, we arrive at the identity
\begin{equation}\label{critim2}
2\ln T_{\rm c}-\frac{\pi z_{0}^2}{2x_{0}}\,T_{\rm c}=\ln\frac{x_{0}^2z_{0}^2}{2(x_{0}^2+z_{0}^2)^3},
\end{equation} 
which can be solved through the Lambert $W$ function:
\begin{equation}
T_{\rm c}=-\frac{4x_{0}}{\pi z_{0}^2}\ W_{-1}\biggl(-\frac{\pi z_{0}^3}{4\sqrt{2}(x_{0}^2+z_{0}^2)^{\frac{3}{2}}}\biggr).
\label{Lambert}
\end{equation}
Index $-1$ of $W$ marks the ${w\leq-1}$ branch of solutions to the equation ${we^w=a}$ in the interval ${-\frac{1}{e}\leq a<0}$.
The solutions exist only if
\begin{equation}
\frac{x_0}{z_0}\geq\sqrt{\left(\frac{e\pi}{4\sqrt{2}}\right)^{\frac{2}{3}}\!\!-1}\equiv\xi\doteq 0.562.
\label{limra}
\end{equation}
The value of $T_{\rm c}$ grows to infinity if ${x_0\to\infty}$ and/or ${z_0\to 0}$, and in this case we can use an approximation ${W_{-1}(a)\approx\ln(-a)-\ln\bigl(-\ln(-a)\bigr)}$.
For the limiting ratio ${x_0/z_0=\xi}$ from Eq.\,\eqref{limra} we get ${T_{\rm c}=4\xi^2/(\pi x_0)}=4\xi/(\pi z_0)$, which for any fixed $x_0$ or $z_0$ represents a minimal value of the crossover time.
If ${x_0/z_0<\xi}$, no sharp crossover time $T_{\rm c}$ can be defined and the transition to the asymptotic regime has a gradual character.

\begin{figure}[tp]
	\includegraphics[width=0.75\linewidth]{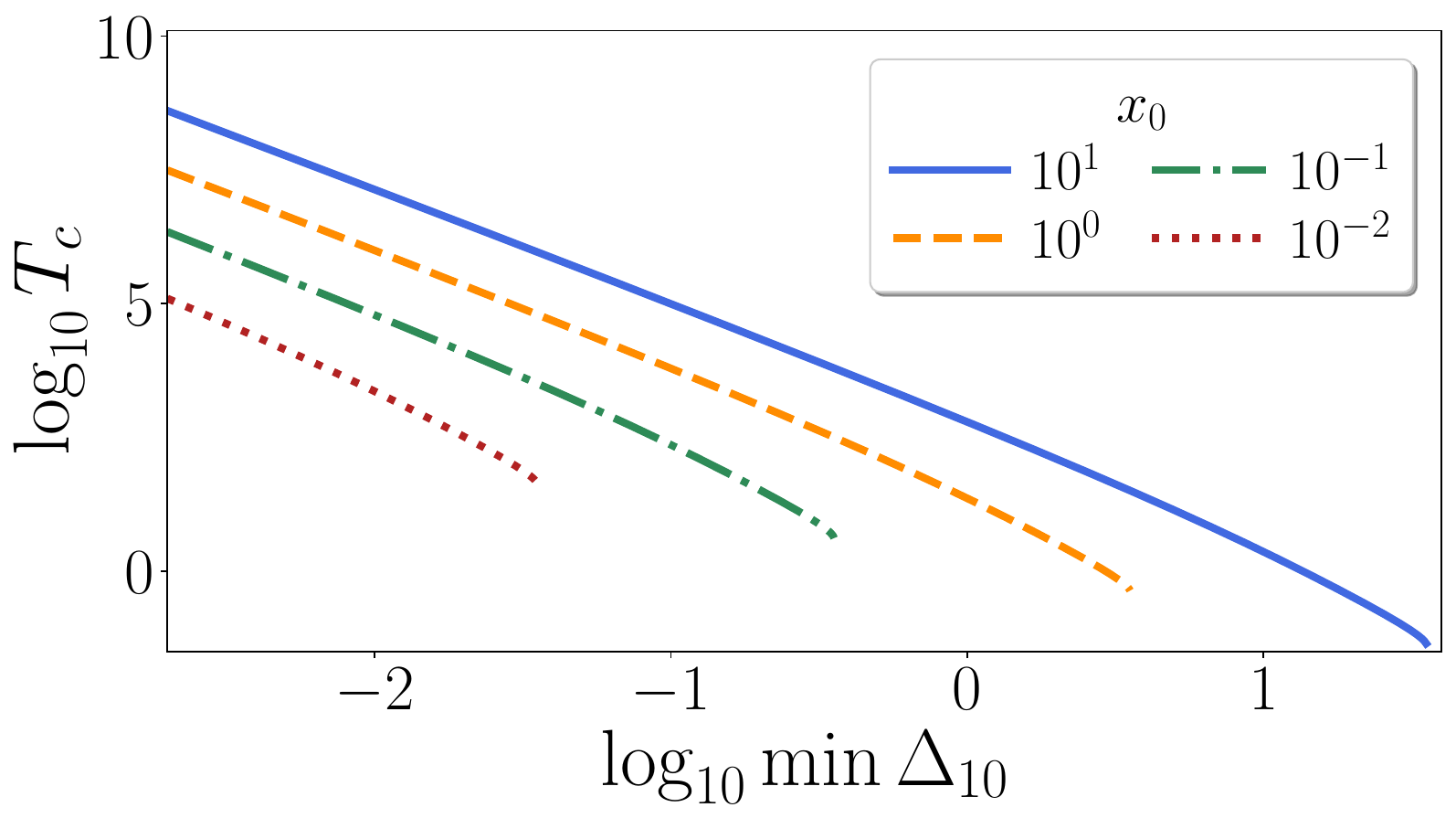}
	\caption{
The crossover time from Eq.\,\eqref{Lambert} as a function of the minimal energy gap, ${{\rm min\,}\Delta(\tau)=2z_0}$, in the two-level model for linear driving protocols~A. 
Individual curves correspond to the indicated values of $x_0$.
Low-$T_{\rm c}$ endpoints of these curves correspond to the saturation of condition \eqref{limra}.  
	}
	\label{fig:Lamb}
\end{figure}

The crossover times predicted by Eq.\,\eqref{Lambert} are shown in Fig.\,\ref{fig:Lamb}.
It depicts the log-log dependence of $T_{\rm c}$ on the minimal energy gap $2z_0$ for various values of $x_0$.
The limiting values following from the condition \eqref{limra} correspond to endpoints of individual curves.
The predicted values of $T_{\rm c}$ are marked by vertical lines in the infidelity dependencies shown earlier in Fig.\,\ref{fig:infLZ}.
We see that formula \eqref{Lambert} works very well.
The infidelity dependence demonstrating the absence of a sharp crossover time for driving parameters violating the condition \eqref{limra} is shown in the inset of Fig.\,\ref{fig:infLZ}.

The polynomial driving protocols discussed in Sec.\,\ref{sub:apliAPT} lead to linear dependencies of $\log{\cal I}(1)_{\rm B}$ (where ${\wp={\rm B}}$ stands for the polynomial paths from Fig.\,\ref{fig:dri}) on $\log T$ with higher slopes.
This was shown in Fig.\,\ref{fig:poly}. 
In principle the transition to these nongeneric asymptotic time regimes is again connected with some crossover times analogous to the above~$T_{\rm c}$.
However, to apply a similar approach as in Eq.\,\eqref{critim2} to the polynomial driving protocols faces two problems:
First, the higher-order coefficients $b_1^{(p)}(1)_{\rm B}$ are difficult to calculate even in the present two-level system.
Second, the Landau-Zener formula would have to be replaced by a more sophisticated expression valid for protocols with ${\dot{s}(\tau)\neq{\rm const}}$.
The dependencies in Fig.\,\ref{fig:poly} indicate that the onset of the asymptotic time regime for polynomial driving protocols happens at times that increase with the degree of the polynomial.
So these protocols do really an excellent job for very long driving times but are not as good for shorter times.

\begin{figure}[tp]
	\includegraphics[width=\linewidth]{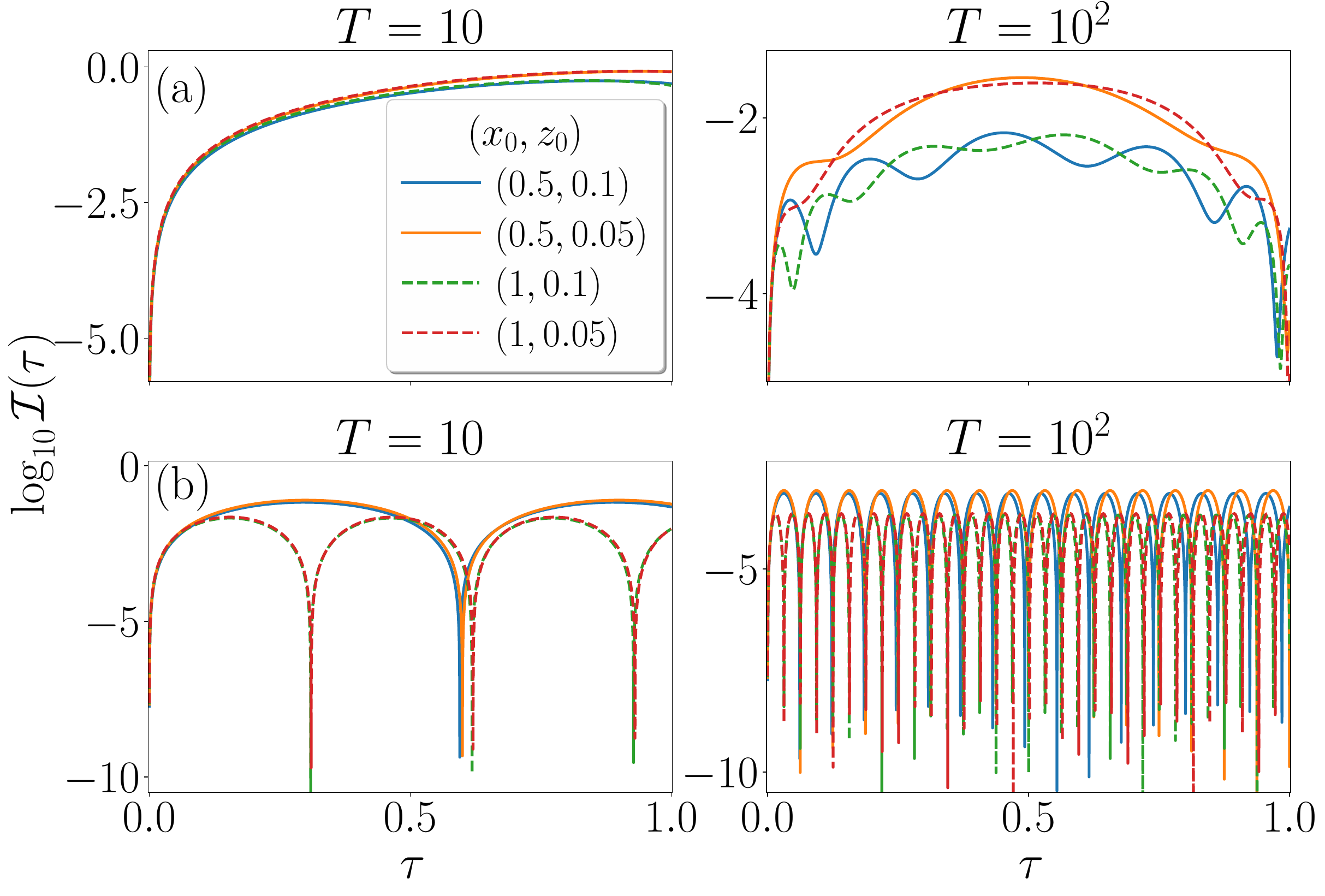}
	\caption{
(Color online) Instantaneous infidelity at time $\tau$ for the geometry-inspired driving protocols in the two-level model: row (a) corresponds to the driving with a constant geodesic speed along line \eqref{eq:li} (protocol~C), row (b) to the ${\dot{s}={\rm const}}$ driving along arc \eqref{eq:arc} (protocol~D).
Columns correspond to two indicated values of~$T$.
Individual curves are assigned to the same parameters as in Fig.\,\ref{fig:infti1}.
Curves with ${x_0 = 0.5}$ and~1 are plotted in solid and dashed linestyle, respectively. 
In row (a) curves with ${z_0 = 0.05}$ lie above those with ${z_0 = 0.1}$ for both choices of $x_0$, in row (b) curves with the same $x_0$ practically overlap. 
	}
	\label{fig:infgeot}
\end{figure}

What about the geometry-inspired types of driving from Sec.\,\ref{se:ge}?
In Sec.\,\ref{Twolev} we explained that geodesics are undetermined in the present model, or in other words, for any pair of parameter points there exists an infinite number of geodesic curves.
Considering, as in the cases above, $(-x_{\rm I},z_{\rm I})=(x_{\rm F},z_{\rm F})=(x_0,z_0)$ with ${x_0,z_0>0}$, we select the following two particular geodesic paths: 
(a) the line \eqref{eq:li} and (b) the arc 
\begin{eqnarray}\label{eq:arc}
\bigl(x(s),z(s)\bigr)&=&r_0\bigl(\cos\alpha(s),\sin\alpha(s)\bigr),
\\
\alpha(s)&=&\alpha_{0}s\!+\!(\pi\!-\!\alpha_{0})(1\!-\!s),\quad s\in[0,1],
\nonumber
\end{eqnarray}
where ${r_0=\sqrt{x_0^2+z_0^2}}$ and ${\alpha_{0}={\rm arctan}(z_0/x_0)}$.
The request of a constant speed \eqref{spee} leads to the conditions $\dot{s}(\tau)\propto r(\tau)^2$ in case (a) and $\dot{s}(\tau)={\rm const}$ in case (b), see Eq.\,\eqref{vconst}.
So the non-geodetic speed $\sqrt{\dot{x}^2\!+\!\dot{z}^2}$ in the $x\times z$ plane varies in case~(a) while it is constant in case~(b). 
Both drivings are of type D from Fig.\,\ref{fig:dri}, but the driving along the line is simultaneously of type C.
So we set ${\wp={\rm C}}$ in case (a) and ${\wp={\rm D}}$ in case (b).

\begin{figure}[tp]
	\includegraphics[width=\linewidth]{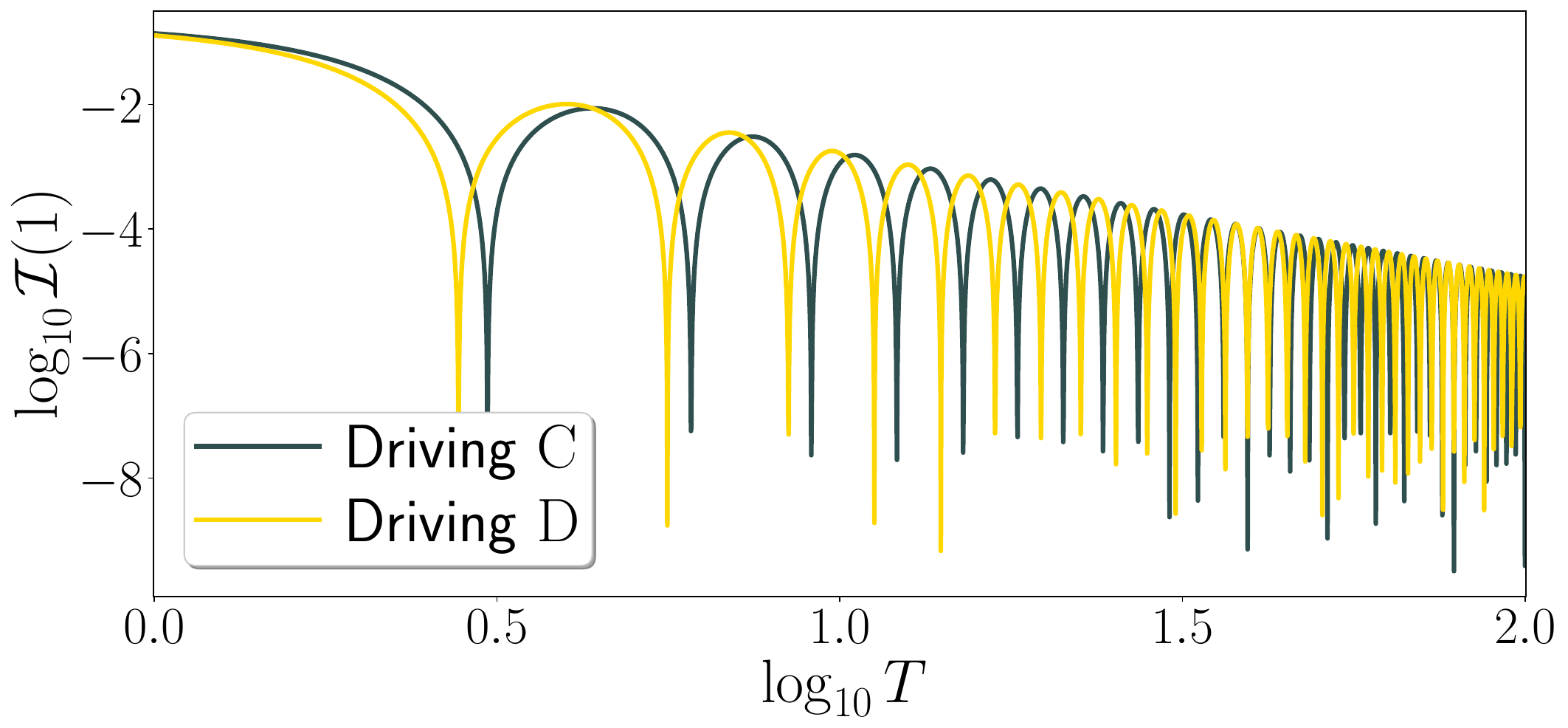}
	\caption{
(Color online) Final infidelity for two geometry-inspired driving protocols~C and~D in the two-level model as a function of the total driving time~$T$. 
Parameters ${(x_0,z_0)=(0.5,1)}$ are the same as in Fig.\,\ref{fig:poly}.
The sharp dips can be proven to reach the value ${{\cal I}(1)_{\wp}=0}$.
	}
	\label{fig:infgeoT}
\end{figure}

Dependencies of the instantaneous infidelity for both geometry-inspired driving protocols are shown in Fig.\,\ref{fig:infgeot}.
Although individual curves in this figure correspond to the same initial and final points as those in Fig.\,\ref{fig:infti1}, their shapes  are qualitatively different.
Across the whole range of time $\tau$ we observe (except the upper left panel) oscillations whose frequency increases and upper boundary decreases with $T$.
For driving along the line, see row~(a), the passage through the avoided-crossing region at ${\tau=0.5}$ is still visible in the dependence ${\cal I}(\tau)_{\rm C}$ for larger final times (here ${T=100}$).
For driving along the arc, see row~(b), the spacing $\Delta(\tau)$ is constant and ${\cal I}(\tau)_{\rm D}$ exhibits periodic dips reaching exact zeros of infidelity at some sharp instants of time (this conclusion, indirectly indicated by numerical results of Fig.\,\ref{fig:infgeot}, can be proven from an analytic solution available for this particular driving). 

\begin{figure}[tp]
	\includegraphics[width=\linewidth]{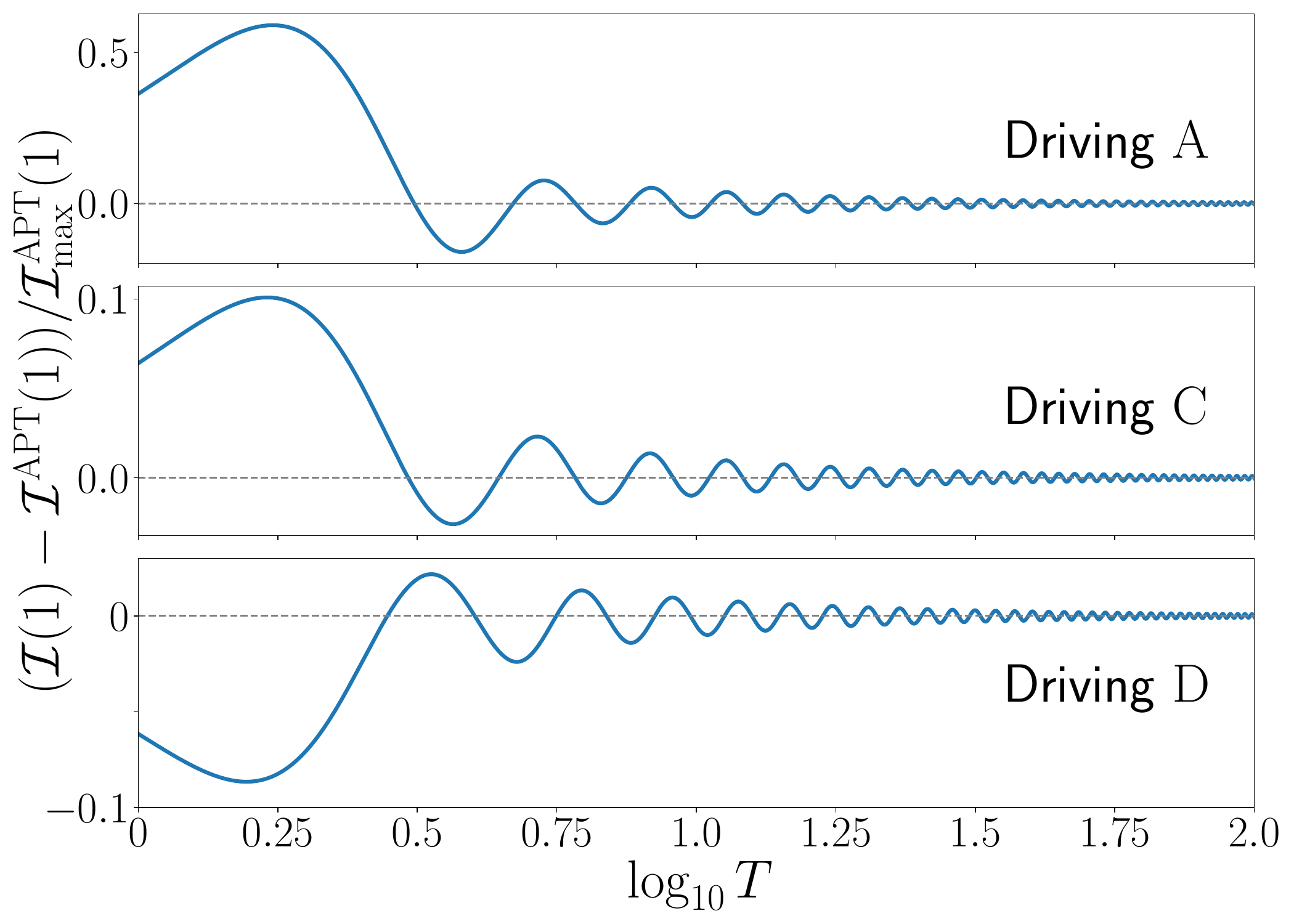}
	\caption{ 
Relative error of the APT approximation in the two-level model for driving protocols A, C and D (the upper, middle and lower panel, respectively) with $(x_0,z_0)=(0.5,1)$.
The $T$-dependent difference between the exact infidelity and the second-order APT infidelity is normalized to the smoothly evolving upper envelope of the APT infidelity.
	}
	\label{fig:exapt1}
\end{figure}

The final infidelity for both geometry-inspired driving protocols is shown in Fig.\,\ref{fig:infgeoT}.
As expected (see Sec.\,\ref{sub:apliAPT}), upper envelopes of both ${\cal I}(1)_{\rm C}$ and ${\cal I}(1)_{\rm D}$ follow a linear log-log decrease with~$T$, the slope taking the predicted value~$-2$.
The upper envelope of both curves is exactly the same, which follows from the expression
\begin{equation}
\bigl|b^{(1)}_1(1)_{\wp}\bigr|^2=\frac{\arctan^2\frac{x_0}{z_0}}{x_{0}^2+z_{0}^2}\sin^2\frac{\varphi_{10}(1)_{\wp}}{2}
\label{bcd}
\end{equation}
that can be obtained for both drivings ${\wp={\rm C}}$ and D. 
Oscillations of both curves differ due to different behavior of the phases $\varphi_{10}(1)_{\wp}$. 

Since Fig.\,\ref{fig:infgeoT} shows results for the same initial and final points as Fig.\,\ref{fig:poly}, the efficiency of all driving protocols for these parameters can be compared.
While the final infidelities for protocols~C and~D are about the same (they only differ in oscillations), the infidelities of protocols~A and~B are systematically lower. 
In the interval $T\in[100,1000]$ (not shown in Figs.\,\ref{fig:poly} and~\ref{fig:infgeoT}, but covered by our calculations) the upper envelope of infidelity in protocol A is reduced by a~factor ${\approx 0.74}$ and the overall averages by ${\approx 0.36}$.
For protocols of type~B the difference reaches several orders of magnitude.
We can conclude that for large enough final times the polynomial protocols in our two-level model provide far better results than all other protocols considered here.
Note however that these differences apply to the upper envelopes of all dependencies and not to the local minima of infidelity at some particular values of~$T$.

Finally, let us discuss the quality of the APT approximation in the two-level model.
In Fig.\,\ref{fig:exapt1} we compare the exact infidelity ${\cal I}(1)_{\wp}$ obtained from the numerical solution with the infidelity ${\cal I}^{\rm APT}(1)_{\wp}$ predicted by the leading-order APT expression \eqref{infilog} for driving protocols~A, C and~D (polynomial protocols~B are not included as the corresponding APT calculations require higher-order terms).
The absolute term of Eq.\,\eqref{infilog} is given by Eqs.\,\eqref{critim} (protocol~A) or \eqref{bcd} (protocols~C and~D).
The difference between the exact and APT infidelities is expressed relative to the smoothly evolving maximal APT infidelity ${\cal I}^{\rm APT}_{\rm max}(1)_{\wp}$ obtained by setting ${\varphi_{10}(1)_{\wp}=\pi}$ in formulas~\eqref{critim} and \eqref{bcd} (the upper envelope of the APT curve) and is depicted as a function of $\log T$.
We see that the APT correctly predicts not only the overall decrease of infidelity, but also its local oscillations.
The relative error of the APT approximation for all three driving protocols is very small already at ${T\approx 10}$ and quickly decreases with increasing~$T$.

\section{Driving in an interacting multi-qubit system}
\label{se:many}

In this Section we proceed from a single qubit to a~system of several mutually interacting qubits.
We use a~specific version of the familiar Lipkin-Meshkov-Glick (or simply Lipkin) model~\cite{Lipk65}.
This model was originally introduced in the context of nuclear physics, but today it often serves as a general example of quantum criticality in a numerically treatable and experimentally realizable many-body system (see, e.g., Refs.\,\cite{Gilm78,Vida06,Cejn07,Ribe08,Orus08,Zibo10,Puri17,Stra18,Cerv21}).

\subsection{The model}

Let us have a system of ${N>1}$ qubits enumerated by indices ${i=1,2,\dots,N}$, each of them endowed with the Hilbert space ${{\cal H}^{(i)}={\mathbb C}^2}$ supporting an algebra of operators $\{\hat{I}^{(i)},\hat{\sigma}_x^{(i)},\hat{\sigma}_y^{(i)},\hat{\sigma}_z^{(i)}\}$ (unit operator and the triple of Pauli matrices).
In the total Hilbert space ${{\cal H}=\otimes_{i=1}^{N}{\cal H}^{(i)}}$ of the full system (dimension ${2^N}$) we introduce so-called quasispin operators
\begin{equation}
\hat{J}_{\bullet}=\frac{1}{2}\sum_{i=1}^{N}\hat{\sigma}_{\bullet}^{(i)},
\qquad{{\bullet}=x,y,z},
\label{J}
\end{equation}
which satisfy commutation relations of angular momentum.
The quasispin algebra conserves the total squared quasispin ${\hat{J}^2\equiv\hat{J}^2_x+\hat{J}^2_y+\hat{J}^2_z} $, so one can restrict the solution to a ${(2j\!+\!1)}$-dimensional subspace ${\cal H}_j$ of ${\cal H}$ characterized by a single value of the total angular momentum quantum number~$j$.
In the following, we set $j$ to the maximal value ${j=\frac{N}{2}}$, whose unique subspace ${\cal H}_j$ with dimension ${d=N\!+\!1}$ is fully symmetric under the exchange of qubits.
 
A Hamiltonian written in terms of quasispin operators \eqref{J} and their products describes a fully-connected system of interacting qubits.
Indeed, any linear combination of quasispin operators can be interpreted as a~one-body Hamiltonian characterizing the total energy of qubits in an external field, while a product of ${n=2,3,\dots}$ quasispin operators represents an $n$-body interaction acting between all qubits of the set.
Here we use a Hamiltonian of the following form,
\begin{eqnarray}
\hat{H}=\ \hat{J}_z-\frac{1}{N}\biggl\{\lambda\hat{J}_x^2&+&\chi\left[\hat{J}_x\bigl(\hat{J}_z\!+\!\tfrac{N}{2}\bigr)+\bigl(\hat{J}_z\!+\!\tfrac{N}{2}\bigr)\hat{J}_x\right]
\nonumber\\
&+&\chi^2\bigl(\hat{J}_z\!+\!\tfrac{N}{2}\bigr)^2\biggr\},
\label{eq:H_lipkin}
\end{eqnarray}
where $\lambda,\chi\in(-\infty,+\infty)$ are two control parameters.
Parameter $\lambda$ represents a relative strength of two-body interactions conserving parity $\hat{\Pi}=(-1)^{\hat{J}_z+N/2}$, while parameter $\chi$ controls parity-violating interactions. 
Here we consider both these parameters as externally controllable.

The phase transitional structure of the model in the parameter plane $\lambda\times\chi$ becomes explicit in the infinite-size limit ${N\to\infty}$.
It can be more intuitively deduced from the bosonic form of Hamiltonian \eqref{eq:H_lipkin}, namely
\begin{eqnarray}
\hat{H}\xrightarrow[]{N\to\infty}&&\ -\frac{N}{2}+\hat{t}^{\dag}\hat{t}
-\frac{1}{N}\biggl[\frac{\lambda}{4}\bigl(\hat{t}^{\dag}\hat{t}^{\dag}\hat{s}\hat{s}+\hat{s}^{\dag}\hat{s}^{\dag}\hat{t}\hat{t}+2\hat{t}^{\dag}\hat{s}^{\dag}\hat{s}\hat{t}\bigr)
\nonumber\\
&&
+\chi\bigl(\hat{t}^{\dag}\hat{t}^{\dag}\hat{s}\hat{t}+\hat{t}^{\dag}\hat{s}^{\dag}\hat{t}\hat{t}\bigr)+\chi^2\bigl(\hat{t}^{\dag}\hat{t}^{\dag}\hat{t}\hat{t}\bigr)\biggr],
\label{Schwing}
\end{eqnarray}
where we use the Schwinger mapping
\begin{equation}
{\bigl(\hat{J}_x\!+\!i\hat{J}_y,\hat{J}_x\!-\!i\hat{J}_y,\hat{J}_{z}\bigr)}\mapsto{\bigl(\hat{t}^{\dag}\hat{s},\hat{s}^{\dag}\hat{t},\tfrac{1}{2}(\hat{t}^{\dag}\hat{t}\!-\!\hat{s}^{\dag}\hat{s})\bigr)}
\end{equation}
of quasispin operators to expressions containing creation and annihilation operators $\hat{s}^{\dag},\hat{t}^{\dag}$ and $\hat{s},\hat{t}$ of structureless bosons of two types: $s$ with positive parity and $t$ with negative parity. 
Thus the form \eqref{Schwing}, which keeps ${\cal O}(N)$ terms and neglects ${\cal O}(1)$ ones, recasts our qubit system as a system of $N$ interacting bosons.   
The phase-transitional analysis has been already presented for a large number of such systems (see, e.g., Refs.\,\cite{Gilm78,Vida06,Cejn07,Ribe08}), so we outline here only the main results.  

\begin{figure}[tp]
	\includegraphics[width=\linewidth]{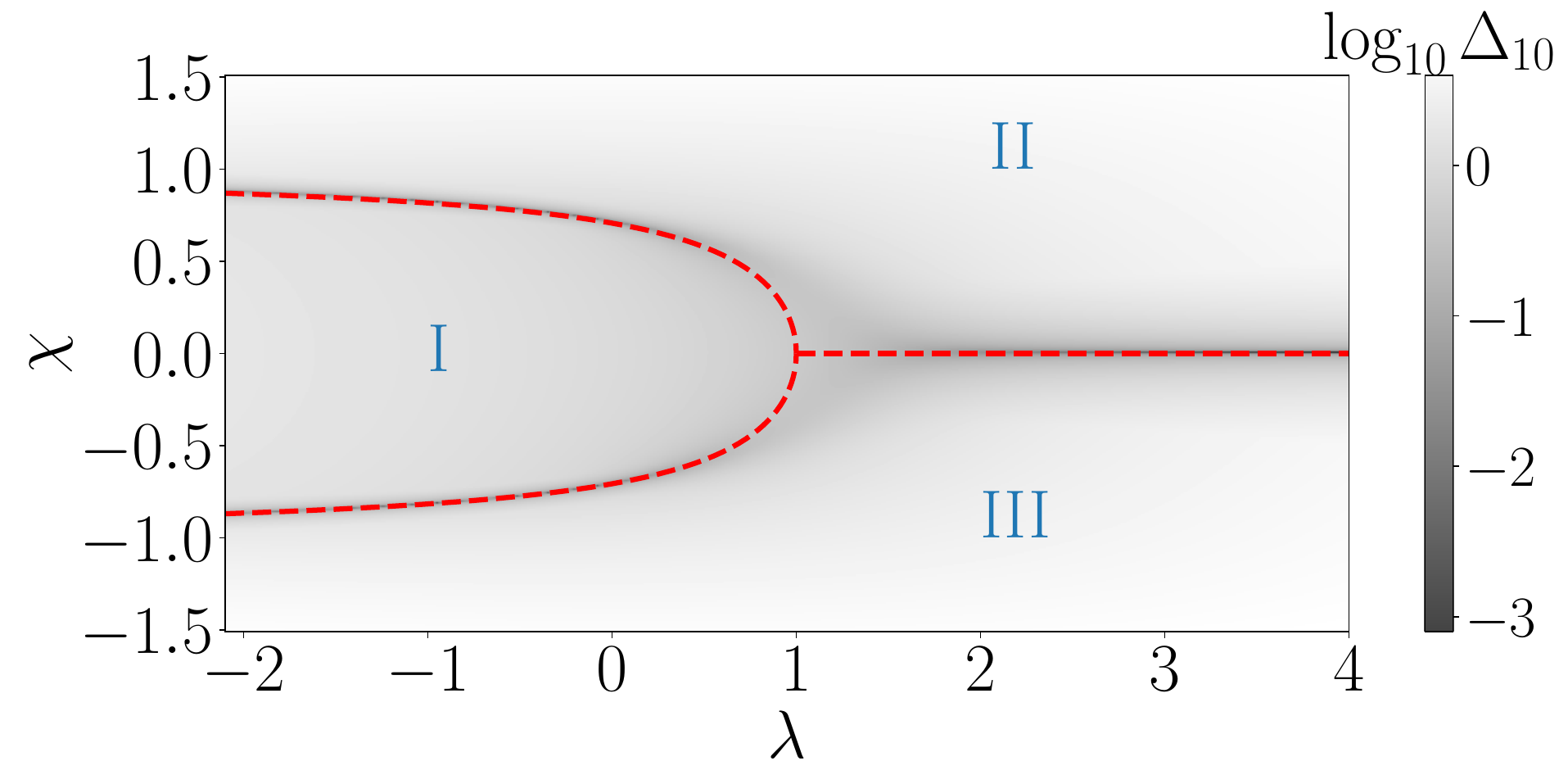}
	\caption{
The phase diagram of Lipkin Hamiltonian \eqref{eq:H_lipkin} in the plane of control parameters.
The dashed curves correspond to the ${N\to\infty}$ ground-state QPTs between phases {I}, {II}, and {III} explained in the text.
The grayscale of the background expresses the size of the energy gap $\Delta_{10}$ for ${N=10}$.
	}
	\label{fig:phas}
\end{figure}

The ground-state of Hamiltonians \eqref{eq:H_lipkin} and \eqref{Schwing}  exhibits QPTs of various types.
For small values of interaction strengths $\lambda$ and $\chi$, the system is in phase~I with the ground-state wave function characterized by expectation values ${\ave{\hat{t}^{\dag}\hat{t}\,}_0=0}$ and ${\ave{\hat{J}_z}_0=-\frac{N}{2}}$.
For $\lambda$ or $\chi$ increasing across a certain critical borderline, the ground state in the ${N\to\infty}$ limit flips to the form with ${\ave{\hat{t}^{\dag}\hat{t}\,}_0>0}$ and ${\ave{\hat{J}_z}_0>-\frac{N}{2}}$.
There are two kinds of this interacting phase: phase~{II} with ${\ave{\hat{t}^{\dag}\hat{s}\!+\!\hat{s}^{\dag}\hat{t}\,}_0>0}$ and $\ave{\hat{J}_x}_0>0$ for ${\chi>0}$, and phase~{III} with ${\ave{\hat{t}^{\dag}\hat{s}\!+\!\hat{s}^{\dag}\hat{t}\,}_0<0}$ and $\ave{\hat{J}_x}_0<0$ for ${\chi<0}$.
The transition between these mirror-symmetric phases also has a critical character. 

The ground-state phase diagram in the plane ${\lambda\times\chi}$ is depicted in Fig.\,\ref{fig:phas} together with a finite-size precursor of criticality---the energy gap $\Delta_{10}$ between the ground state and the first excited state for ${N=10}$.
The gap at the critical borderlines vanishes in the infinite-size limit, and this happens exponentially (${\Delta_{10}\propto e^{-aN}}$, where ${a>0}$ is a constant) in the first-order phase transition, or algebraically (${\Delta_{10}\propto N^{-p}}$, where ${p>0}$ is a rational power) in the second-order phase transition.
In the phase diagram of Fig.\,\ref{fig:phas}, all ground-state QPTs are of the first order, except the \uvo{triple point} $(\lambda,\chi)=(1,0)$, where the phase transition is of the second order.

The above-described phase structure of the model strongly affects geometric properties of its ground-state manifold.
The metric tensor can be calculated numerically and, for finite qubit numbers $N$, it is nonsingular, except some isolated diabolic points in the parameter plane where the gap $\Delta_{10}$ accidentally vanishes (we will show elsewhere that these points appear on the finite-$N$ precursor of the QPT separatrix in the ${\lambda<0}$ half-plane).
However, in the limit ${N\to\infty}$, the gap is zero everywhere on the QPT separatrices, which implies divergence of the metric tensor and impassability of the separatrices for geodesic curves~\cite{Kuma12}.
Since we study only finite-$N$ systems, the QPT-induced singularities of the metric structure are not actually present.
They are virtual, showing up only through precursors of the infinite-size behavior.
Let us note that the $N$-dependent geometric structure of the ground-state manifold of Hamiltonian \eqref{eq:H_lipkin} is rather complex and will be analyzed in a separate paper.

\subsection{Results and discussion}
\label{se:re2}

The ${j=\frac{N}{2}}$ ground state of the Lipkin Hamiltonian at any parameter point $\bLambda$ can be expanded in the eigenbasis $\ket{m}$ of the $\hat{J}_z$ operator, 
\begin{equation}
\ket{E_0(\bLambda)}=\sum_{m=-N/2}^{+N/2}a_m(\bLambda)\,\ket{m}, 
\end{equation}
with $a_m(\bLambda)$ denoting normalized amplitudes.
The initial parameter point for all driving paths is chosen as ${\bLambi\equiv(\lambda_{\rm I},\chi_{\rm I})=(0,0)}$, where the ground state  reads ${\ket{E_0(\bLambi)}=\ket{m=-\frac{N}{2}}}$.
This is a totally uncorrelated state of qubits expressed as ${\ket{0}_1\otimes\ket{0}_2\otimes\cdots\otimes\ket{0}_N}$.
The target state, i.e., the ground state at a selected final parameter point $\bLambf$, has a more complex structure.
Let us stress that within the present model the target state is always classically computable in a polynomial time with respect to $N$, so we use it here merely to benchmark the performance of various driving protocols.
The final point $\bLambf$ is chosen at various places of the ${\lambda\times\chi}$ plane, either still in phase~I or in phase~{II}. 
Let us note that phase~{III} does not need to be considered because of its formal equivalence to phase~{II} (this means that any path in the ${\chi>0}$ half-plane has its identical mirror-symmetric image in the ${\chi<0}$ half-plane).
Crossing of the QPT separatrix implies that the target state $\ket{E_0(\bLambf)}$ is spread in the $\ket{m}$ basis and contains strong correlations between individual qubits induced by their mutual interactions. 
This conforms with the idea of adiabatic quantum computation, but simultaneously induces the problem of passing the critical parameter region where the energy gap $\Delta_{10}$ becomes infinitely small as the number of qubits $N$ asymptotically increases.

\begin{figure}[tp]	
	\includegraphics[width=\linewidth]{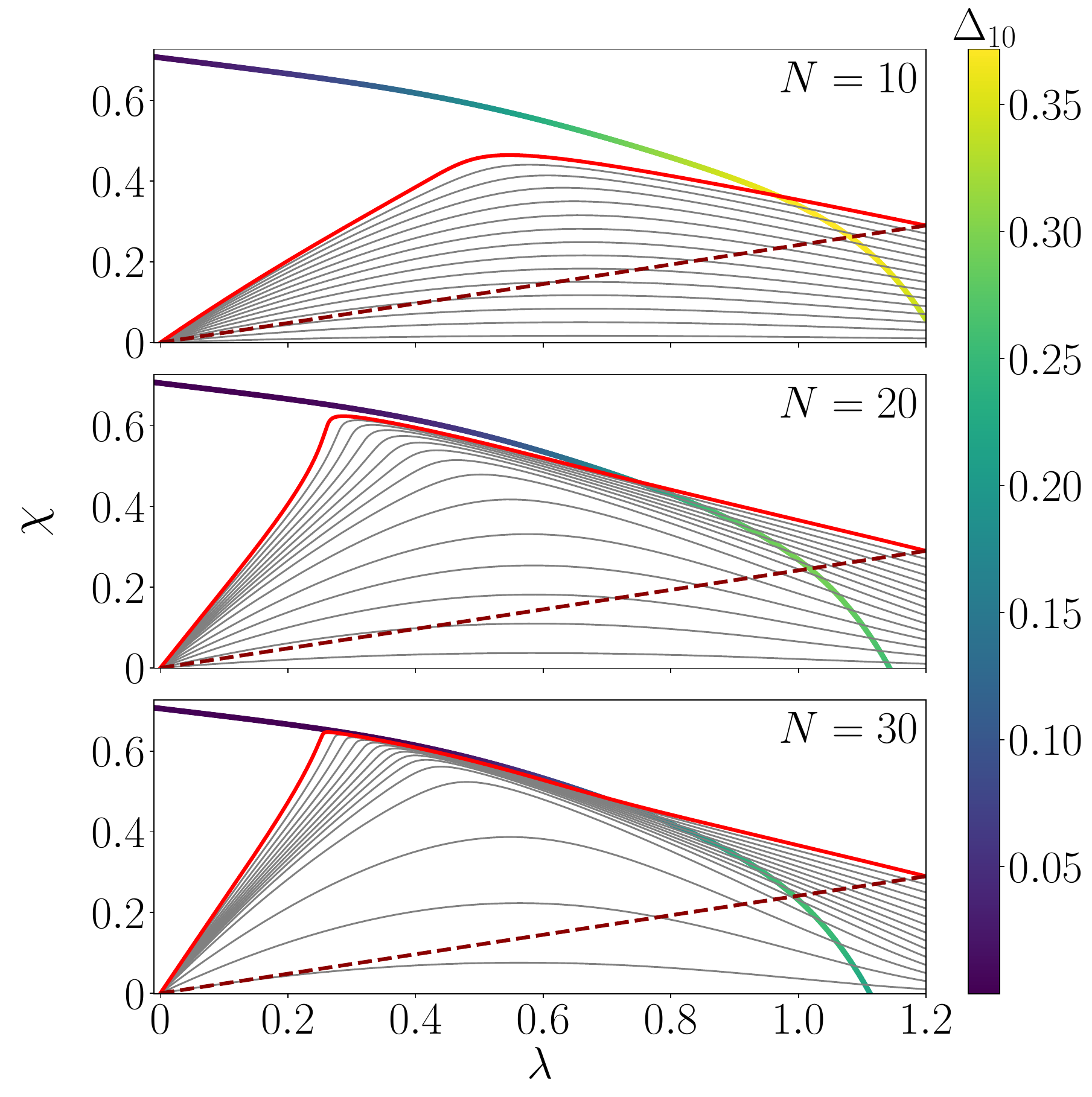}
	\caption{
(Color online) Geodesic curves of the Lipkin model \eqref{eq:H_lipkin} connecting point ${(\lambda_{\rm I},\chi_{\rm I})}={(0,0)}$ with ${(\lambda_{\rm F},\chi_{\rm F})}={(1.2,i/100)}$, where ${i=1,3,\dots,29}$, for different qubit numbers $N$.
The precritical curves of  the minimal energy gap $\Delta_{10}$ are depicted in each panel associated with a given $N$, the values of $\Delta_{10}$ being indicated by varying color along these curves.
The geodesic to the last point $(1.2,0.29)$ is highlighted (red online) and the linear path to the same point is also shown (dashed line).
	}
	\label{fig:geodesics_lipkin}
\end{figure}

We will again compare all driving protocols from Sec.\,\ref{sec:drivings} (Fig.\,\ref{fig:dri}).
To do so, we first calculate the metric tensor~\eqref{metr} on the ground-state manifold and then determine the geodesic paths for various final parameter points~$\bLambf$.
The latter is done with the aid of Eq.\,\eqref{geoeq}, which is solved as a boundary-value problem for the function $\Lambda^{\mu}(s)$ (where we further set ${s=\tau}$) with Dirichlet boundary condition ${\Lambda^{\mu}(0)=\Lambi^{\mu}}$ and ${\Lambda^{\mu}(1)=\Lambf^{\mu}}$.
As mentioned above, the metric tensor of the present ${N>1}$ model is nonsingular (except isolated diabolic points) and the geodesics nondegenerate.
So the driving path along the geodesic (protocol D) and those along the line (protocols A, B and C) have different geometric lengths.

Figure~\ref{fig:geodesics_lipkin} depicts geodesic curves connecting the initial point ${(\lambda_{\rm I},\chi_{\rm I})=(0,0)}$ with different final points ${(\lambda_{\rm F},\chi_{\rm F})}$ for several values of the qubit number~$N$.
Each panel, associated with a given $N$, also shows a precritical curve, i.e., the curve demarcating the minimal energy gap $\Delta_{10}$ (a finite-$N$ version of the QPT separatrix).
The value of~$\Delta_{10}$ is encoded into the indicated color scale.
We see that for ${\chi_{\rm F}\neq 0}$ the driving trajectories always cross the precritical curve in the region of the first-order QPT.
We even observe attraction of the geodesic curves to the parts of the precritical curve with smaller values of the gap.
Although this may seem counterintuitive (one could guess that the geodesics will try to avoid parameter domains with large values of the metric tensor), the displayed behavior represents true solutions to Eq.\,\eqref{geoeq} (we invoke an analogy with gravitational attraction to spatial regions containing large masses).   
As a consequence, numerical determination of geodesics becomes a challenge for large numbers of qubits since $\Delta_{10}$ drops exponentially with~$N$ at the first-order QPT.

\begin{figure}[tp]
	\includegraphics[width=\linewidth]{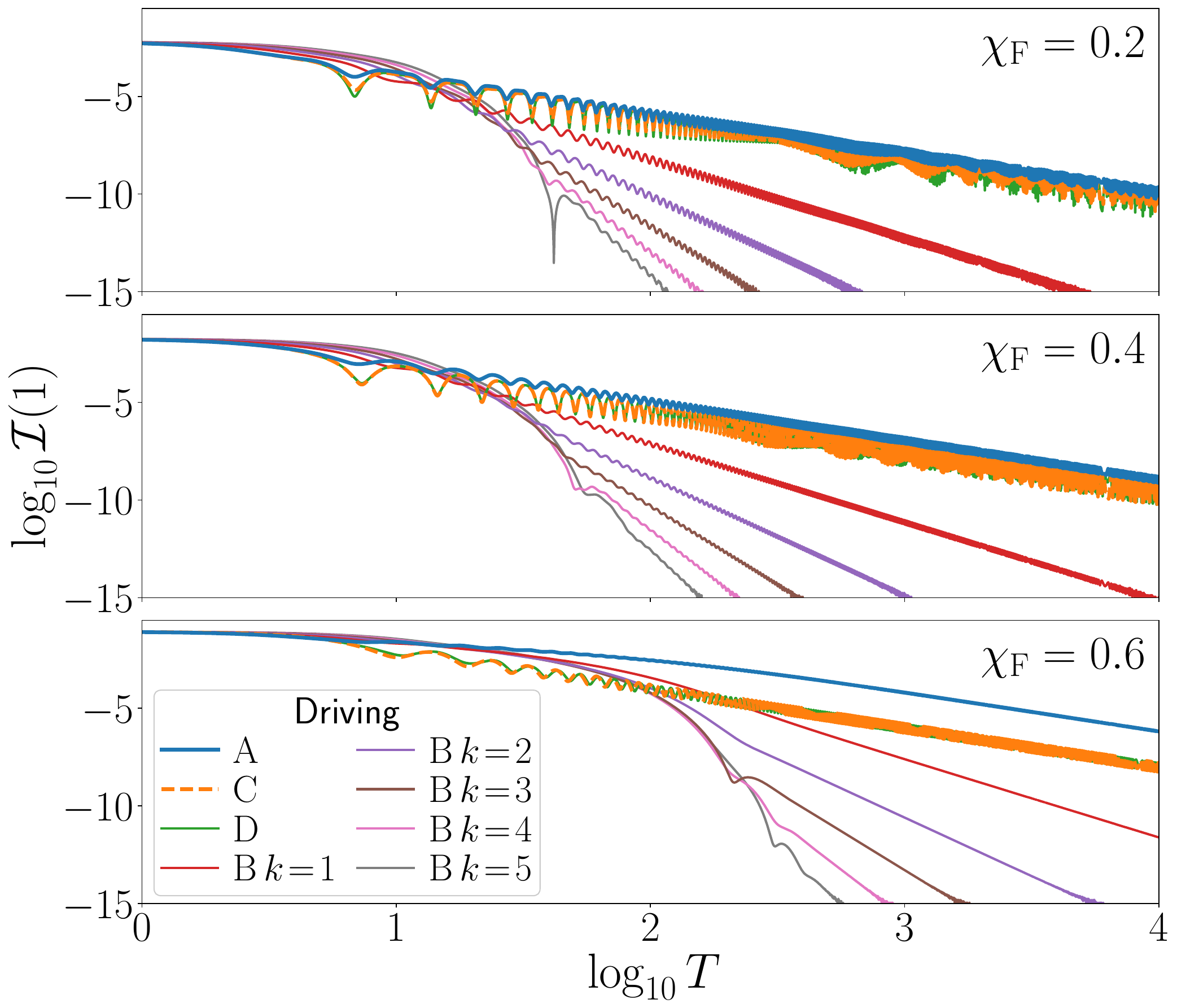}
	\caption{
(Color online) A log-log plot of the final infidelity ${\cal I}(1)_{\wp}$ as a~function of the total driving time $T$ in the Lipkin model with ${N=10}$ for various driving protocols whose trajectories do not cross the QPT separatrix.
The curves corresponding to various protocols are distinguished by color (the ordering of curves for large $T$ is the same as the ordering of the legend, protocols C and D are distinguished by dashed and solid lines but mostly overlap).
The trajectories lead from ${(\lambda,\chi)=(0,0)}$ to ${(0.3,\chi_{\rm F})}$, with $\chi_{\rm F}$ specified in each panel.
	}
	\label{fig:Lipkin_no_crossing}
\end{figure}

\begin{figure}[tp]
	\includegraphics[width=\linewidth]{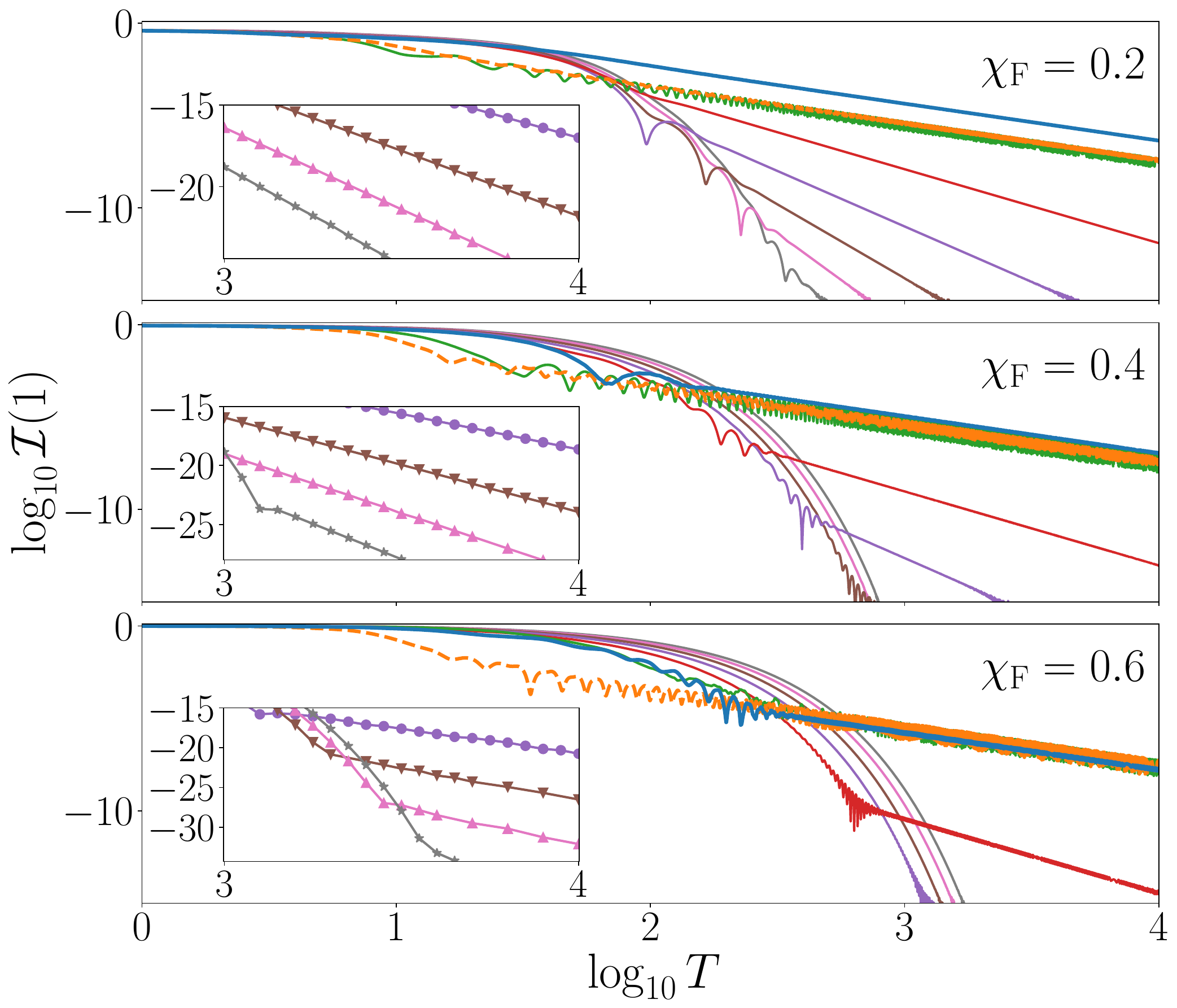}
	\caption{
(Color online) The same as in Fig.\,\ref{fig:Lipkin_no_crossing}, but for driving protocols whose trajectories end at ${(\lambda,\chi)=(1.2,\chi_{\rm F})}$ behind the QPT separatrix (the line colors distinguish the same protocol types).
The parts of the dependencies with lower infidelity are shown with a lower time resolution in the insets; symbols $\bullet$, {\scriptsize$\blacktriangledown$}, {\scriptsize$\blacktriangle$} and {\normalsize$\star$} correspond to polynomial protocols with $k =2, 3, 4$ and 5, respectively.  
	}
	\label{fig:Lipkin_crossing}
\end{figure}

Calculations of the final infidelity for all driving protocols A, B (${k=1,\dots,5}$), C and D in the Lipkin model with ${N=10}$ are presented in Figs.\,\ref{fig:Lipkin_no_crossing} and \ref{fig:Lipkin_crossing}.
The driving trajectories lead from ${(\lambda_{\rm I},\chi_{\rm I})=(0,0)}$ to various final points ${(\lambda_{\rm F},\chi_{\rm F})}$.
Figure~\ref{fig:Lipkin_no_crossing} shows results for the trajectories that do not cross the QPT separatrix (${\lambda_{\rm F}=0.3}$, final point in phase~I) and Fig.\,\ref{fig:Lipkin_crossing} collects results for the trajectories that cross the QPT separatrix (${\lambda_{\rm F}=1.2}$, final point in phase~II).
While all curves in Fig.\,\ref{fig:Lipkin_no_crossing} pass through the crossover to the asymptotic regime within the displayed infidelity range ${{\cal I}(1)_{\wp}>10^{-15}}$, some of the curves in Fig.\,\ref{fig:Lipkin_crossing} reach the crossover for ${{\cal I}(1)_{\wp}<10^{-15}}$.
That is why the low-fidelity parts of the dependencies are shown in the insets of Fig.\,\ref{fig:Lipkin_crossing}.
Because of larger computational demands implied by the required high accuracy, the low-fidelity parts of the dependencies were calculated with a lower resolution on the time axis.

Considering all the main panels and insets together, we can say that the dependencies in Figs.\,\ref{fig:Lipkin_no_crossing} and~\ref{fig:Lipkin_crossing} manifest qualitatively the same features as those in the two-level model (cf.\,Figs.\,\ref{fig:poly}, \ref{fig:infLZ} and \ref{fig:infgeoT}). 
At very short driving times~$T$, the infidelity for all protocols starts from nearly the same value, depending only on positions of the initial and final parameter points in the given one- or multi-qubit system (exact convergence of all curves would be observed at ${T=0}$).
The explanation follows from the sudden approximation, which equates the short-time fidelity with the overlap of the ground-state eigenvectors at the initial and final points.
We notice that the overlap is larger (the infidelity smaller) if the initial and final points lie in the same quantum phase of the system (Fig.\,\ref{fig:Lipkin_no_crossing}) than if they lie in different phases (Fig.\,\ref{fig:Lipkin_crossing}).
In the domain of medium driving times~$T$, the infidelity dependencies exhibit roughly an exponential overall decrease connected with smaller or larger oscillations.
In this domain, the ordering of individual curves quickly varies with $T$ and depends also on the choice of the final parameter point.
This evolution lasts until the curves---each one at a different time---reach the asymptotic regime.  
In the domain of very long driving times~$T$, we observe just a~linear log-log decrease (still accompanied by some oscillations) with the slope directly deduced from the APT. 

Both Figs.\,\ref{fig:Lipkin_no_crossing} and \ref{fig:Lipkin_crossing} clearly demonstrate that for very long driving times the final infidelity is fairly the best (smallest) for the polynomial driving protocols~B.
This dominance increases with time $T$ and with the degree of the polynomial as the slopes of the corresponding graphs increase with~$k$. 
In this domain, the linear driving protocol~A as well as the geometry-inspired protocols~C and~D yield much worse results depending on~$(\lambda_{\rm F},\chi_{\rm F})$.
On the other hand, polynomial protocols~B reach the asymptotic regime later than the others, the delay being proportional to $k$ and also depending on~$(\lambda_{\rm F},\chi_{\rm F})$.
This leads to a better performance of protocols~A, C and~D in some time windows at smaller values of $T$.
The advantage of these protocols at medium times is larger for driving trajectories across the QPT separatrix (Fig.\,\ref{fig:Lipkin_crossing}) than for those confined within the same phase (Fig.\,\ref{fig:Lipkin_no_crossing}).
Nevertheless, there is no clear winner of this competition since the optimal medium-time protocol depends on~$(\lambda_{\rm F},\chi_{\rm F})$ and is sensitive to~$T$.
We can observe that for the driving trajectories leading to larger values of $\chi_{\rm F}$ (${\geq 0.4}$) across the QPT separatrix (Fig.\,\ref{fig:Lipkin_crossing}), the medium-time performance of protocol~C becomes systematically better than that of both protocols~D and~A.
We assume that the disadvantage of protocol~D follows from its inclination towards the small-gap domain near the QPT separatrix (see Fig.\,\ref{fig:geodesics_lipkin}).
In this domain, $\dot{\bLambda}(\tau)$ is very small, and therefore, big values of $\ddot{\bLambda}(\tau)$ are needed in the rest of the trajectory to keep the fixed total time $T$, which delays the onset of the asymptotic regime.

\begin{figure}[tp]
	\includegraphics[width=0.75\linewidth]{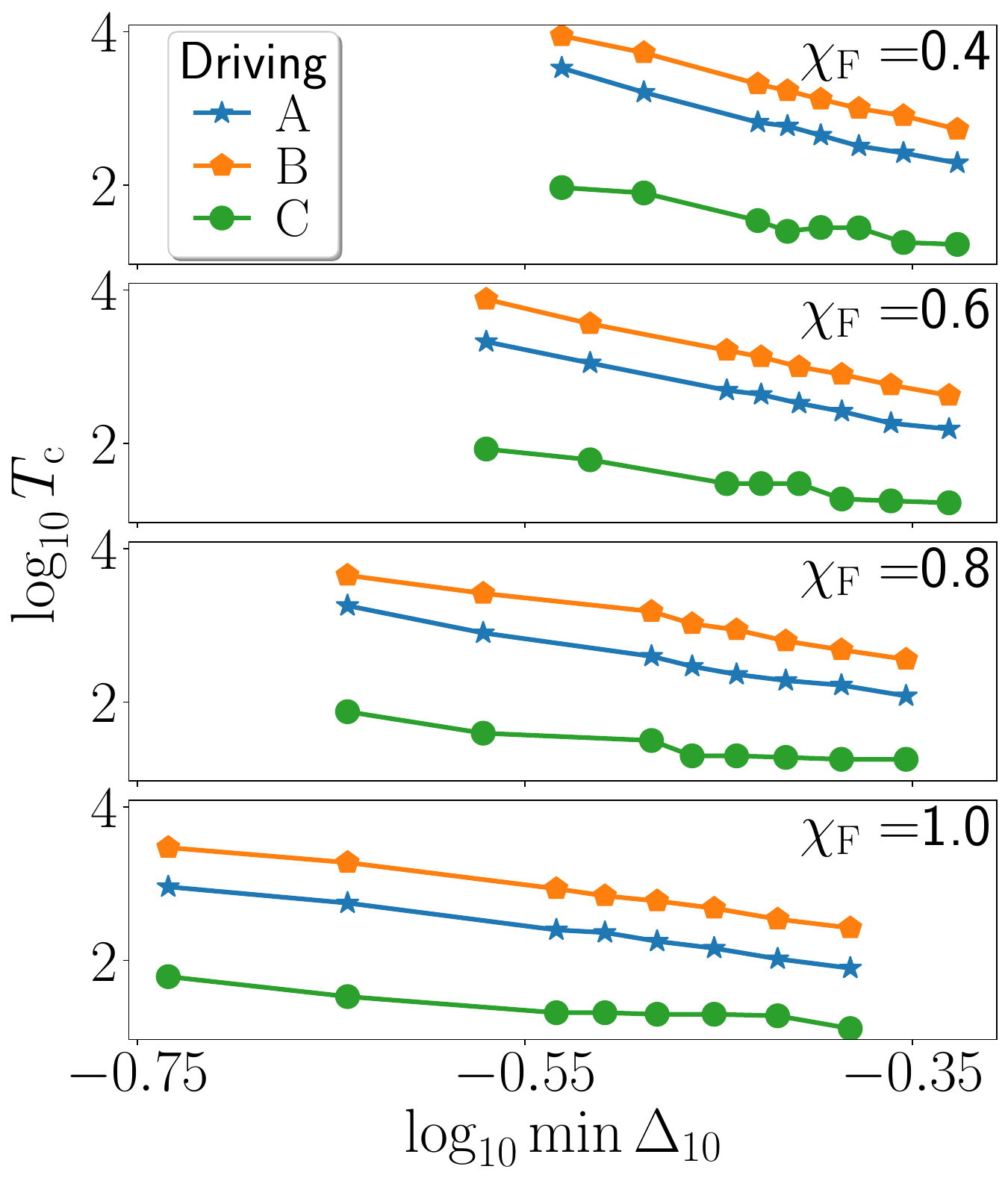}
	\caption{
Log-log dependencies of the crossover time $T_{\rm c}$ on the minimal energy gap $\Delta_{10}$ for driving protocols A, B (${k=1}$) and C in the Lipkin model.
The driving trajectory is a line from ${(\lambda,\chi)=(0,0)}$ to $(1.5,\chi_{\rm F})$, with $\chi_{\rm F}$ specified in each panel.
The minimal energy gap is associated with crossing of the QPT separatrix and is reduced by increasing the number of qubits (the points on the graphs correspond to ${N = 5}$,\,6,\,7,\,8,\,9,\,10,\,15,\,20).
	}
	\label{fig:Critical_final_times}
\end{figure}

As mentioned in the above discussion, an important role in selecting an optimal driving protocol is played by the crossover time $T_{\rm c}$ from the preasymptotic regime of driving (characterized by an approximately exponential decrease of the final infidelity with the driving time~$T$) to the asymptotic regime (characterized by an algebraic decrease of infidelity following from the APT).
Although no analytic expression can be derived for the crossover time in the present multi-qubit system, approximate values of~$T_{\rm c}$ can be determined from graphs of the final infidelity for different qubit numbers~$N$.
The results are shown in Fig.\ref{fig:Critical_final_times}.
It presents log-log dependencies of the crossover time $T_{\rm c}$ on the minimal energy gap $\Delta_{10}(\tau)$ along the line from $(\lambda_{\rm I},\chi_{\rm I})=(0,0)$ to final points $(\lambda_{\rm F},\chi_{\rm F})$ across the QPT separatrix (${\lambda_{\rm F}=1.5}$) for various protocols along the given line: the linear driving~A, polynomial diving~B with ${k=1}$, and the constant-speed driving~C. 
The variation of the minimal gap is achieved via changing the size parameter $N$. 
We observe a qualitatively similar dependence (a~roughly algebraic decrease) of $T_{\rm c}$ on ${\rm min\,}\Delta_{10}(\tau)$ as in the two-level model (Fig.\,\ref{fig:Lamb}), where we however analyzed only the linear driving.

Figure~\ref{fig:Critical_final_times} manifests a clear hierarchy of crossover times for the three driving protocols. 
The smallest values of~$T_{\rm c}$ are systematically achieved for the protocol~C, medium values of~$T_{\rm c}$ characterize the protocol~A, and finally the highest values of~$T_{\rm c}$ are observed for the protocol~B (when the crossover leads to the asymptotic regime with slope $-4$ instead of $-2$). 
Let us note that the crossover times for polynomial protocols with ${k>1}$ are increasingly higher than those for ${k=1}$. 
The crossover for these protocols occurs at decreasing values of infidelity, which would make a detailed evaluation of $T_{\rm c}$ (whose dependence on the minimal gap also shows roughly an algebraic decrease) more time consuming.
The information obtained from Fig.\,\ref{fig:Critical_final_times} extends the results discussed in connection with the medium-time dependencies in Figs.\,\ref{fig:Lipkin_no_crossing} and \ref{fig:Lipkin_crossing}.
It allows us to conclude that a~good driving strategy (the best among those tested here) in the medium-time domain, at least if the initial and final points lie in different quantum phases of the system, is the one inscribed in the geometry-inspired protocol~C.
It combines two aspects which turn out to be important: (a) a linear trajectory across the QPT separatrix, which ensures a larger minimal energy gap than the geodesic path, and (b) the requirement of the constant speed on the manifold, which reduces the losses of fidelity when crossing the minimal gap domain and apparently advances the transition to the asymptotic regime.

\begin{figure}[tp]
	\includegraphics[width=\linewidth]{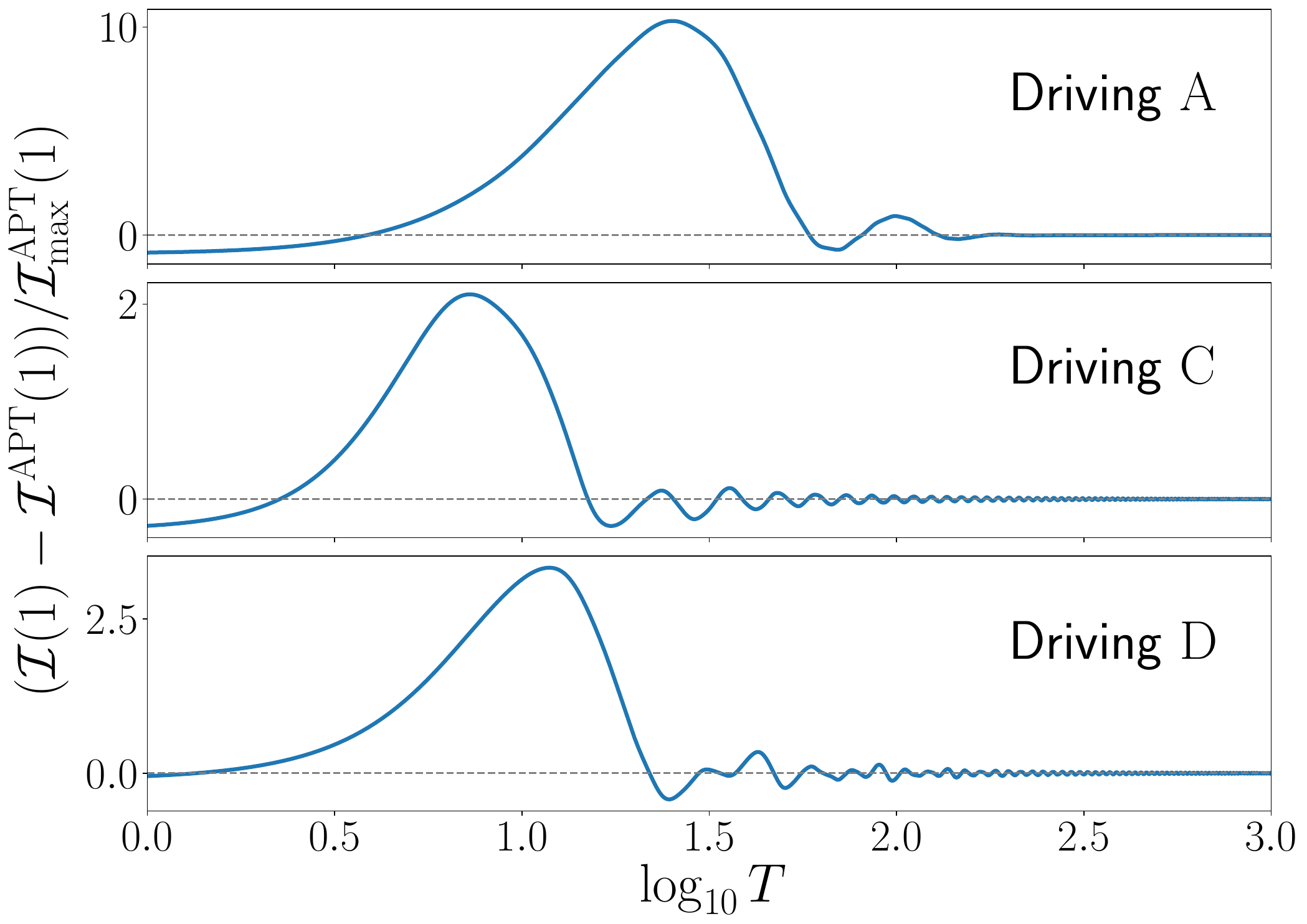}
	\caption{
Relative error of the APT approximation in the Lipkin model with ${N=10}$ for driving protocols A, C and D from ${(\lambda,\chi)=(0,0)}$ to $(1.2,0.4)$.
The meaning of symbols is the same as in Fig.\,\ref{fig:exapt1}.
	}
	\label{fig:exapt2}
\end{figure}

As in the one-qubit system (Sec.\,\ref{se:two}), the results obtained in the present multi-qubit model manifest that the driving protocol~D using exact geodesics in the parameter space is suboptimal in majority of cases covering both medium- and asymptotic-$T$ regimes.
We again invoke the argument based on coherence of quantum dynamics for isolated systems.
Indeed, unitary evolution of coherent quantum superpositions of the immediate Hamiltonian eigenstates allows for nonmonotonous variations of instantaneous fidelity along the driving path.
In such situations, the maximal fidelity at the final point is apparently not guaranteed by the minimal length of the driving trajectory in the sense of Provost--Vallee metric.

Finally, as in Sec.\,\ref{se:re1}, we present a short comparison of the results derived from the APT with those obtained from exact simulations of driven dynamics in the multi-qubit system. 
In Fig.~\ref{fig:exapt2}, the exact final infidelity ${\cal I}(1)_{\wp}$ from numerical simulations for ${N=10}$ is compared with the leading-order (${\propto T^{-2}}$) APT prediction ${\cal I}^{\rm APT}(1)_{\wp}$ for protocols~A, C and~D along a~line crossing the QPT separatrix.
As in Fig.\,\ref{fig:exapt1}, the difference between both infidelities is normalized to the smoothly evolving upper envelope ${\cal I}^{\rm APT}_{\rm max}(1)_{\wp}$ of the APT infidelity.
The comparison is shown within the range ${T\in [1,1000]}$, and we observe that the agreement becomes almost perfect for ${T\gtrsim 300}$.

\section{Conclusion}
\label{CONC}

In this paper, we design and test several driving protocols with the aim to maximize the fidelity of preparation of a~correlated pure state of a~quantum many-body system. 
The initial state and the target state of the driving procedure are supposed to be ground states of the system belonging to different quantum phases, so the driving trajectory in the parameter space has to cross a~finite-size precursor of a QPT.
As a~toy model for preliminary tests of our approaches, we use a single-qubit (two-level) system with a single avoided crossing of levels.
The analysis is then extended to an interacting fully-connected multi-qubit system with several types of QPT.

An essential point of our analysis is the use of the adiabatic perturbation theory of Refs.\,\cite{Orti08,Orti10,Orti14}.
This theory in its leading order is shown to give remarkably good predictions of infidelity for sufficiently long driving times.
Adopting the APT to the optimal driving problem, we find a hierarchy of polynomial driving protocols that maximize the fidelity in the very long time domain---above a~sharp crossover from the medium-time (Landau-Zener) driving regime to the asymptotic time APT regime.
The dominance of these protocols in the asymptotic regime is really strong, but the crossover time to this regime increases roughly in an algebraic way with a decreasing minimal energy gap between the ground-state and first excited state along the driving trajectory.
If the trajectory crosses a finite-size precursor of a QPT separatrix, the minimal gap drops with the size of the system (exponentially for the first-order QPT or algebraically for a continuous QPT).
Therefore, the applicability of the new protocols is hindered by the system size, which sets limits on scalability of quantum state preparation techniques based on these protocols.
This is not surprising, and in principle conforms with similar conclusions discussed previously in connection with the Landau-Zener regime of driving (see, e.g., Ref.\,\cite{Schu06}).

The second important aspect of our study is the use of geometry-inspired driving protocols in systems with higher than one-dimensional parameter space.
The determination of the metric structure of the ground-state manifold in such systems and calculation of geodesic curves is an interesting problem on its own, and we intend to present this analysis in a separate paper. 
However, the driving protocols based on the full solution to the geodesic problem turned out to yield mostly suboptimal results in the situations discussed.
A~better driving strategy in the medium-time domain seems to follow from the idea of keeping a constant speed on the ground-state manifold along an arbitrary (artificially designed) driving trajectory that avoids parameter regions with small energy gap.
Our study is inconclusive in selecting the optimal driving protocol for the medium-time domain, but we clearly disprove the conjecture on general supremacy of the geodesic protocols \cite{Tomk16,Kolo17}.
We believe that this is because of quantum coherence effects that necessarily appear in the dynamics of isolated quantum systems but are not reflected by the Provost--Vallee definition of the metric tensor (interpreted as the infidelity induced by an infinitesimal quench in the parameter space).
Therefore the Berry's question \cite{Berr88} on the physical significance of geodesic trajectories for quantum systems remains open.

\acknowledgements
This work was supported by the Czech Science Foundation under the Grant No. 20-09998S and by the Charles University in Prague under the project UNCE/SCI/013.


 \end{document}